\documentclass[reprint,aps,nofootinbib,superscriptaddress,preprintnumbers]{revtex4-2}
\usepackage{braket}
\usepackage{slashed}
\usepackage{graphicx}
\usepackage{xcolor}
\usepackage{enumitem}
\usepackage{siunitx}
\usepackage{mathtools}
\usepackage{dsfont}
\usepackage[hidelinks]{hyperref}
\usepackage{tikz}
\usepackage{amsmath,amsthm,amsfonts,amscd,amssymb} 
\usepackage{soul}

\usetikzlibrary{patterns}

\newcommand\blfootnote[1]{%
  \begingroup
  \renewcommand\thefootnote{}\footnote{#1}%
  \addtocounter{footnote}{-1}%
  \endgroup
}

\newcommand{\dt}{\frac{d}{dt}}
\newcommand{\angbracket}[1]{\left\langle #1 \right\rangle}
\newcommand{\mb}{\mathbf}
\newcommand{\x}{\mathbf{x}}

\newcommand{\mbs}[1]{\boldsymbol{#1}}

\DeclareMathOperator{\csch}{csch}
\DeclareMathOperator{\sech}{sech}

\begin{document}

\title{Minimal noise in non-quantized gravity}
\author{Giuseppe Fabiano}
\affiliation{Physics Division, Lawrence Berkeley National Laboratory, Berkeley, CA}
\author{Tomohiro Fujita}
\affiliation{Department of Physics, Ochanomizu University, Bunkyo, Tokyo 112-8610, Japan}
\affiliation{Kavli Institute for the Physics and Mathematics of the Universe (Kavli IPMU),
WPI, UTIAS, The University of Tokyo, Kashiwa, Chiba 277-8568, Japan}
\author{Akira Matsumura}
\affiliation{Department of Physics, Kyushu University, Fukuoka 819-0395, Japan}
\affiliation{Quantum and Spacetime Research Institute, Kyushu University, Fukuoka 819-0395, Japan}
\author{Daniel Carney}
\affiliation{Physics Division, Lawrence Berkeley National Laboratory, Berkeley, CA}

\begin{abstract}
An elementary prediction of the quantization of the gravitational field is that the Newtonian interaction can entangle pairs of massive objects. Conversely, in models of gravity in which the field is \emph{not} quantized, the gravitational interaction necessarily comes with some level of noise, i.e., non-reversibility. Here, we give a systematic classification of all possible such models consistent with the basic requirements that the non-relativistic limit is Galilean invariant and reproduces the Newtonian interaction on average. We demonstrate that for any such model to be non-entangling, a quantifiable, minimal amount of noise must be injected into any experimental system. Thus, measuring gravitating systems at noise levels below this threshold would be equivalent to demonstrating that Newtonian gravity is entangling. As concrete examples, we analyze our general predictions in a number of experimental setups, and test it on the classical-quantum gravity models of Oppenheim et al., as well as on a recent model of Newtonian gravity as an entropic force.
\end{abstract}

\maketitle

\tableofcontents

\blfootnote{gfabiano@lbl.gov, fujita.tomohiro@ocha.ac.jp, \\
matsumura.akira@phys.kyushu-u.ac.jp, carney@lbl.gov}

\section*{Introduction}

At energy densities below the Planck scale $m_{\rm Pl}/\ell_{\rm Pl}^3 \approx 8\times 10^{90}~{\rm g}/{\rm cm^3}$, gravity can be consistently quantized. In this regime, the metric $g_{\mu\nu}$ can be expanded around a background like flat spacetime $g_{\mu\nu} \approx \eta_{\mu\nu} + h_{\mu\nu}$, and the small strain perturbations $h_{\mu\nu}$ are treated as an ordinary quantum field theory. They are quantized into gravitons, just like electromagnetic fields are quantized into photons~\cite{Donoghue:1994dn,Burgess:2003jk,Donoghue:2022eay}. There is currently a large amount of experimental effort aiming to determine if this is the correct description of how gravity operates in nature~\cite{Carney:2018ofe,Lindner:2004bw,Kafri:2013wxa,Bose:2017nin,Marletto:2017kzi,Matsumura:2020law,Datta:2021ywm,Carney:2021yfw,Oppenheim:2022xjr,Lami:2023gmz}.

To make these tests precise, we also need alternative hypotheses: what exactly would it mean for gravity to be ``not quantized'', and how could we detect the difference? In this paper, by ``not quantized'', we mean any model that is \emph{not} the usual graviton model. There are now a swath of concrete examples, including Penrose-Diosi style models~\cite{diosi1987universal,penrose1996gravity,Tilloy:2015zya,tilloy2016sourcing,Giulini:2022wyp}, measurement-feedback models~\cite{Kafri:2014zsa,Carney:2023aab,Piccione:2025ieb,Piccione:2026rzq}, ``classical-quantum'' models~\cite{Oppenheim:2018igd,Layton:2023oud,Oppenheim:2023izn,Kryhin:2023cyw,Carney:2024izr}, and simple toy models of gravity as an emergent entropic force~\cite{Carney:2025wmz}. What has been lacking so far is a general framework that can parametrize all of these models, and a clear set of statements about what experiments can distinguish between these models and perturbatively quantized gravity. 

In this paper, we provide such a general framework, and use it to propose concrete experimental targets. In particular, we show that any model in which gravity is not quantized into gravitons predicts a small but irreducible amount of noisy, irreversible evolution in massive objects. This conclusion requires only the minimal assumptions that the model is Galilean invariant, time-local, and reproduces Newton's laws on average. Thus, if an experiment could be done which shows that masses evolve with less noise than the thresholds reported here, then one can conclude that the gravitational interaction must be entangling. 

Quantitatively, the level of noise corresponding to gravitational entanglement is small but not astronomically smaller than state-of-the-art force measurements. For example, for a pair of masses placed $1~{\rm mm}$ apart, the noise threshold is roughly $\sqrt{S_{aa}} \approx 10^{-18}~{\rm m}/{\rm s^2}/\sqrt{\rm Hz}$, about a factor of 1000 below the LISA Pathfinder level~\cite{armano2024depth}. Moreover, this means that as experiments probe increasingly lower levels of noise, they will quantitatively exclude larger portions of the parameter space of the non-graviton models. For a concrete example, we show the parameter space of classical-quantum gravity in Fig.~\ref{fig:cq}.

Bounds on the noise injected in non-entangling gravitational models and their relation to entanglement generation have appeared before, notably in the pioneering work of Kafri and Taylor~\cite{Kafri:2013wxa}, and more recently by a number of authors in a variety of contexts~\cite{Oppenheim:2022xjr,Lami:2023gmz,Angeli:2025ojs}. Our treatment should be viewed as an extension of these works. A major motivation for the work reported here was the discovery that the earlier calculations cannot accurately capture the noise in some of the concrete models constructed in recent years, as discussed below. We believe that the framework presented here, which is based on just the minimal assumptions of Galilean invariance and averaged Newtonian evolution, captures the full space of possible ``non-quantized'' gravitational models in the non-relativistic limit.

\section{Parametrization of gravitational models}
\label{sec:model construction}

We begin by providing a general form of the time evolution for massive objects under gravitational interactions. This will take the form of a Lindblad master equation. It can be applied to both perturbatively quantized gravity as well as all currently known models of non-quantized gravity, including those cited in the introduction. The equations of motion have a handful of free parameters that can be matched to any particular model. Our treatment will be in the non-relativistic limit for the masses, since this is where most experimental effort is currently focused. The expectation is that a relativistic model of gravity will reduce to our master equation in the non-relativistic limit. 

We give a number of explicit examples of all of these statements in Sec.~\ref{sec:examples}. There we show how our framework can be used to describe the ``classical-quantum'' models of Oppenheim et al.~\cite{Oppenheim:2018igd,Layton:2023oud,Oppenheim:2023izn} and the entropic gravity model of~\cite{Carney:2025wmz}. These examples also serve to highlight some features of our general framework which were missed in earlier treatments of this problem, namely non-Gaussian evolution on individual masses and correlated gravitational noise acting on pairs of masses.

To construct the time evolution rules, we make the following three physical assumptions: \\

\noindent \textbf{1. Newton's law of gravity holds on average (``Ehrenfest theorem'')}: The Heisenberg equations of motion in expectation value coincide with the classical Hamilton equations for two masses interacting gravitationally, namely
    \begin{equation}
    \label{eq:classicallimit}
    \begin{aligned}
        &\dt \angbracket{\mb x_{a}}=\frac{\angbracket{\mb p_{a}}}{m_{a}} \, , \\ &\dt\angbracket{\mb p_{a}}=\angbracket{\frac{d}{d\mb x_{a}} \sum_{b \neq a} \frac{G_N m_a m_b }{|\mb x_a-\x_b|}} \, .
        \end{aligned}
    \end{equation}
This is a minimal condition which means that the gravitational dynamics of large objects, such as planets and stars, obey the usual Newtonian laws. In particular, for any object with sufficiently small variance in position and momentum, classical gravity still holds. \\

\noindent \textbf{2. Time locality}: The evolution law is local in time. This is of course true in ordinary Newtonian gravity, in either classical or quantum mechanics. In certain non-graviton models of gravity, there is some degree of non-locality in time, because there are unobserved ``baths'' which can lead to non-Markovian features in the evolution of the observed massive objects. Concretely, ``classical-quantum'' (CQ) gravity has this property, in principle~\cite{Oppenheim:2018igd,Layton:2023oud,Oppenheim:2023izn}. However, the non-relativistic limit of this model is time-local to very good approximation~\cite{Layton:2023oud,Carney:2024izr}. All other models we are aware of also obey time locality explicitly. 

Mathematically, time locality implies that we can model the evolution of massive objects with a Lindblad equation
\begin{equation}
\label{eq:startinglindblad}
    \dot{\rho} =\mathcal{L}[\rho]=-i[H,\rho]+\mathcal{D}(\rho) \, ,
\end{equation}

where $H$ is the Hamiltonian of the system and the non-unitary or dissipative part of the evolution takes the general form 
\begin{equation}
\label{eq:dissipationinitial}
\begin{aligned}
    \mathcal{D}(\rho)=\int d\lambda \,\left[ L_\lambda \rho L_\lambda^\dagger - \frac{1}{2}\{L_\lambda^\dagger L_\lambda,\rho\}\right] \, ,
    \end{aligned}
\end{equation}
where $\lambda$ is a general index that can be discrete and/or continuous. We will assume that the dissipator in the non-relativistic limit is a function of the positions only $L_{\lambda} = L_{\lambda}(\mb{x}_1, \mb{x}_2, \ldots)$, for simplicity and because this holds in all known example models, but it is easy to relax this condition if desired.

In terms of the Lindbladian $\mathcal{L}$, the Ehrenfest theorem conditions become
\begin{equation}
    \begin{aligned}
    \label{eq:channel_class_lim}
        &\mathrm{Tr}[\mathcal{L}(\rho)\mb x_a]=\frac{\angbracket{\mb p_{a}}}{m_{a}} \, , \\ &\mathrm{Tr}[\mathcal{L}(\rho)\mb p_a]=\angbracket{\frac{d}{d\mb x_{a}} \sum_{b \neq a} \frac{G_N m_a m_b }{|\mb x_a-\x_b|}} \, .
        \end{aligned}
    \end{equation}
This means that average Newtonian dynamics may be realized in various ways. In the majority of the non-relativistic models studied in the literature ~\cite{diosi1987universal,penrose1996gravity,Tilloy:2015zya,tilloy2016sourcing,Giulini:2022wyp,Oppenheim:2018igd,Layton:2023oud,Oppenheim:2023izn,Kryhin:2023cyw,Carney:2024izr,Carney:2025wmz}, the Newton potential appears in the Hamiltonian, so we have
\begin{equation}
\label{eq:ham_initial}
\begin{aligned}
    &H=\sum_{a}\frac{\mb p_a^2}{2m_a}- \sum_{a < b} \frac{G_N m_a m_b}{|\mb x_a-\mb x_b|} , \\
    &\mathrm{Tr}[\mathcal{D}(\rho)\mb x_a]=\mathrm{Tr}[\mathcal{D}(\rho)\mb p_a]=0 \, .
    \end{aligned}
\end{equation}
Another possibility is that the Hamiltonian only contains the kinetic energy and the Newtonian interaction arises from the non-unitary part of the dynamics:
\begin{equation}
\label{eq:ham_initial_2}
\begin{aligned}
    &H=\sum_{a}\frac{\mb p_a^2}{2m_a} , \qquad 
    \mathrm{Tr}[\mathcal{D}(\rho)\mb x_a]=0 \, ,\\
    &\mathrm{Tr}[\mathcal{D}(\rho)\mb p_a]=\angbracket{\frac{d}{d\mb x_{a}} \sum_{b \neq a} \frac{G_N m_a m_b }{|\mb x_a-\x_b|}} \, .
    \end{aligned}
\end{equation}
For examples of these types of models, see \cite{Carney:2023aab,Piccione:2025ieb,Piccione:2026rzq}. By linearity, one could also engineer $\mathcal{L}(\rho)$ such that the average Newtonian evolution arises partly from the Hamiltonian and partly from the dissipation terms, although we do not know of any explicit examples like this. \\

\noindent \textbf{3. Galilean symmetry}: The dynamics is invariant under the action of the standard non-relativistic symmetries, i.e., the Galilean group. This means under a Galilean transformation we have
    \begin{equation}
    \label{eq:symmetrycondition}
       \dot\rho=\mathcal{L}[\rho]\, \rightarrow \dot\rho'=\mathcal{L}[\rho'] \, ,
    \end{equation}
where $\rho'=U\rho U^\dagger$ and the unitary $U$ implements an arbitrary Galilean transformation. \\

These assumptions are enough to fix the evolution law into a simple parametrized form. The Hamiltonian term takes the form \eqref{eq:ham_initial} or \eqref{eq:ham_initial_2} plus any other one- or two-body operators describing, e.g., internal energies and non-gravitational interactions, which we will omit for brevity. The more complex part is the dissipation $\mathcal{D}$. A detailed computation (Appendix \ref{app:symmetryapp}) shows under our assumptions, the most general Lindblad operators for a pair of masses $\mb{x}_1, \mb{x}_2$ are reduced to the form
\begin{equation}
L_{\lambda} = L_{\mb{k},i} = e^{i\mathbf{k}\cdot\mathbf{R}}J_{\mathbf{k},i}(\mathbf{r})
\end{equation}
where $\mb{r} = \mb{x}_1 - \mb{x}_2$ and $\mb{R} = (m_1 \mb{x}_1 + m_2 \mb{x}_2)/(m_1 + m_2)$ are the relative and center of mass coordinates respectively. All remaining model dependence is contained in the rotationally invariant functions $J_{\mathbf{k},i}(\mathbf{r})$. Explicitly,
\begin{equation}
\label{eq:diss_main}
\begin{aligned}
    \mathcal{D}(\rho)=&\sum_i    \int d^3\mb k \Big[e^{i\mb k\cdot \mb R}J_{\mb k,i}(\mb r)\rho \,  J^\dagger _{\mb k,i}(\mb r)e^{-i\mb k\cdot \mb R}\\
    &-\frac{1}{2}\{J^\dagger_{\mb k,i}(\mb r)J_{\mb k,i}(\mb r),\rho\}\Big] \, .
    \end{aligned}
\end{equation}
The $J_{\mb{k},i}$ are not totally arbitary. The Ehrenfest conditions Eq.~\eqref{eq:channel_class_lim} constrain them in specific ways depending on whether the Newtonian interaction comes from a Hamiltonian or dissipative term, as outlined in Appendix \ref{app:model_restriction}. We will use the specific forms of these constraints below. To keep the presentation streamlined, in the remainder of the main text we focus primarily on models in which the Newtonian interaction is included in the Hamiltonian, as this class encompasses most of the models studied in the literature. A parallel derivation for the case of a purely dissipative Newtonian interaction is provided in Appendix \ref{app:two_oscillators}, leading to identical noise bounds.

Equation~\eqref{eq:diss_main} is one of the central results of this paper. It gives the most general possible evolution of two non-relativistic massive objects undergoing mutual gravitational interactions. In perturbatively quantized gravity, the $J_{\mb k,i} \equiv 0$, and time evolution is reversible, up to negligibly small corrections due to emitted gravitons. Any non-zero dissipation terms represent a deviation from standard graviton physics, and in particular irreversibility of time evolution.\footnote{Of course, we are also ignoring any real experimental noise sources in Eq.~\eqref{eq:diss_main}. The dissipation term here only includes effects due to the gravitational interactions. In the rest of the paper, when we talk about measuring noise rates due to such gravitational effects, what we really mean is doing a a search for anomalously large noise. We are only going to be interested in placing \emph{upper bounds} on the noise. So, if an experiment is done where the \emph{total} actual noise is measured to be, say, $\Gamma_{\rm meas}$, then one can safely conclude that the gravitational noise $\Gamma_{\rm grav} \leq \Gamma_{\rm meas}$.} All such models can be captured by an appropriate choice of these $J_{\mb{k},i}$ functions.

We emphasize that the form of Eq.~\eqref{eq:diss_main} \emph{does not} rely on any Gaussianity assumptions, in contrast to previous approaches to this problem~\cite{Kafri:2013wxa,Oppenheim:2022xjr,Lami:2023gmz,Angeli:2025ojs}. In those works, the equations were further simplified by assuming that the position operators are ``small'', leading to expansions up to second order in $|\mb{x}_a|$. The intuition behind this is that we are thinking of a lab experiment with two masses separated by a distance $d$ (of order, say, millimeter or larger), and the fluctuations of the mass positions $|\mb{x}_a| \ll d$, as shown in Fig.~\ref{fig:cartoons}. However, this does not accurately capture the noise acting on individual masses. Indeed, consider a single massive object by itself in the vacuum of space. Many of the gravitational models considered here cause even this system to decohere. But in this scenario, there is no baseline $d$ to expand $|\mb{x}_a|$ around. Put another way, only interaction terms involving both masses can reliably be expanded in a Gaussian approximation. In the examples in Sec.~\ref{sec:examples}, we will see that it is important to keep track of the non-Gaussianity in the single-mass dissipation terms.

With this in mind, we end by recording a slightly simplified form of Eq.~\eqref{eq:diss_main}, which is capable of describing all of the example models we are aware of. Keeping arbitrary single-particle terms, expanding any two-body terms in $|\mb{r}|/d \ll 1$, and restricting to the spatial axis $x$ defined by the two masses, we obtain
\begin{equation}
\label{eq:lind_restr}
\begin{aligned}
    & \dot{\rho}=-i [H_0, \rho] - i \alpha_G \left[ r^2,\rho \right]
    \\
    & + \sum_{a=1,2}\int d^3\mb k f_a(k)\left(e^{i\mb{k} \cdot \mb{x}_a}\rho e^{-i \mb{k} \cdot \mb{x}_a}-\rho\right)  +\beta [x_1,[\rho,x_2]] \, .
    \end{aligned}
\end{equation}
See Appendix~\ref{app:symmetryapp} for a derivation. The first line gives ordinary unitary evolution, including any non-gravitational terms $H_0$ and the Newtonian entangling interaction expanded to leading order in $|\mb{r}|/d \ll 1$. The gravitational coupling strength at distance $d$,
\begin{equation}
\label{eq:alphaG}
\alpha_G = \frac{G_N m_1 m_2}{d^3},
\end{equation}
will appear repeatedly. The second line describes the dissipation. In particular, the terms proportional to $f_a$ are single-body dissipation terms, and in general can be non-quadratic (i.e., leading to non-Gaussian evolution), depending on the positive and isotropic real functions $f_a(\mb{k})$.\footnote{Earlier papers typically wrote the single-body dissipation in the form $\dot{\rho} \sim \sum_{a=1,2} \alpha_a [\mb{x}_a,[\mb{x}_a,\rho]]$, because they made a Gaussian approximation. In our more general framework, these can be recovered but only if $f_a$ has certain singular terms. In general and in specific models shown below, there are actually \emph{no} such quadratic single-body dissipation terms. See Appendix~\ref{app:model_restriction} for some more details.} The final term, proportional to the real coefficient $\beta$, represents ``non-local'' noise that acts on both masses. All told, this means that we can parametrize the evolution in an arbitrary model with a single real free parameter $\beta$ and some real functions $f_a(\mb{k})$. In a given model, these free parameters depend on the masses $m_a$ and distances $d$ in the problem in specific ways.

\section{Noise bounds and gravitational entanglement}
\label{sec:main}

\begin{figure*}[ht!]

\includegraphics[width=\linewidth]{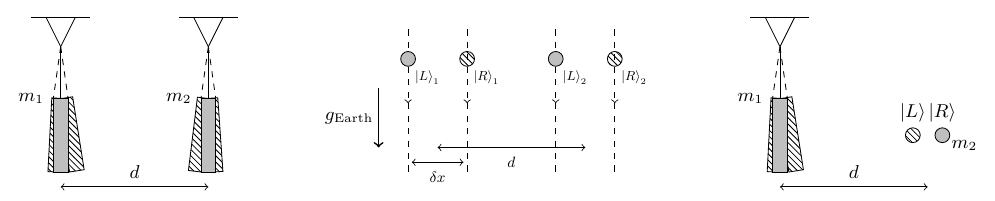}

\caption{\textbf{Schematic experimental setups.} Left: two mechanical oscillators placed at distance $d$, used to measure their mutual gravitational interaction~\cite{Westphal:2020okx,Agafonova:2024evr,Manley:2026evd}. Fluctuations in the force are schematically depicted by the dashed masses. Center: a pair of free-falling masses, approximated as two-state systems $\ket{L,R}$ and initially superposed into two locations, following~\cite{Bose:2017nin}.  Right: a hybrid system consisting of a mechanical source mass coupled to a two-state system, such as an atom interferometer near a large source mass~\cite{kovachy2015quantum,asenbaum2017phase,Carney:2021yfw,overstreet2022observation}.}
\label{fig:cartoons}

\end{figure*}

We now turn to an analysis of noise induced by the non-reversible nature of the non-quantized gravity models, and how it relates to their ability or inability to entangle masses. The competition of these effects can be understood in a simple way from Eq.~\eqref{eq:lind_restr}. The Newtonian interaction generates entanglement, while the single-body dissipation terms destroy it. The non-local noise term proportional to $\beta$ can actually help to alleviate the entanglement loss, a subtle point which we discuss in detail below. Put simply, the only way a model can fail to generate entanglement is if the single-body noise is sufficiently strong.

Mathematically, the strategy to show this is straightforward and follows~\cite{Kafri:2013wxa}. The total noise coming from the single-body dissipation terms is quantified by comparing the average rate of change of some quantity to the rate expected if there was no dissipation. Concretely, for an operator $\mathcal{O}$, let
\begin{equation}
\braket{\mathcal{O}}_e = \braket{\mathcal{O}} - \braket{\mathcal{O}}_H,
\end{equation}
where  
\begin{equation}
\braket{\mathcal{O}}_H = {\rm tr}~\rho_H \mathcal{O}, \ \ \ \rho_H = e^{i H t} \rho(0) e^{-i H t}
\end{equation}
gives the operator evolved in the absence of any dissipation. The subscript ``e'' stands for excess. With this definition we can define noise rates
\begin{equation}
\label{eq:noise_schematic}
\Gamma = \frac{d\Braket{\mathcal{O}}_e}{dt}   \Big|_{t=0}.
\end{equation}
We will always discuss these rates evaluated at $t=0$, although for notational simplicity we generally drop the $t=0$ subscripts. The specific choice of the observables $\mathcal{O}_a$ depend on the systems involved. For mechanical oscillators and free particles, we will use the momentum variance $\mathcal{O}=\mb{p}^2$, while for two-state systems like an atom interferometer, $\mathcal{O}=\sigma_-=\ket{0} \bra{1}$ detects the loss of phase contrast in a superposition. We can then compare the rate of noise generation in a given model to the net rate of entanglement generation under the same evolution. 

We find quite generally there is a threshold value of the noise $\Gamma_{\rm thresh}$ such that if the noise is below this threshold $\Gamma \leq \Gamma_{\rm thresh}$, the gravitational interaction must be entangling. A similar set of bounds were found under more restrictive assumptions in~\cite{Kafri:2013wxa,Angeli:2025ojs}. Proving this needs to be done in a slightly case-by-case fashion depending on the physical systems. We begin with the case of a pair of mechanical masses, before moving on to a variety of other architectures, all shown in Fig.~\ref{fig:cartoons}.

\subsection{Two mechanical oscillators}

We begin with the case of a pair of mechanical systems, for example oscillators with masses $m_{1,2}$ and frequencies $\omega_{1,2}$. This kind of experimental configuration was first proposed in~\cite{Kafri:2013wxa} and is the subject of many current efforts, e.g.~\cite{Westphal:2020okx,Agafonova:2024evr,Manley:2026evd}. See Fig.~\ref{fig:cartoons}, left. 

What we will do is first compute a noise observable for these masses, followed by an entanglement observable, and then show that if the noise is below a certain minimal level, entanglement must be generated. Here we outline the argument and central results, with full calculational details provided in Appendix~\ref{app:two_oscillators}.

To quantify noise, we consider the variance in the momenta. Let
\begin{equation}
\braket{p^2}_e = \braket{p^2} - \braket{p^2}_H
\end{equation}
denote the ``excess'' momentum squared, where again excess means the part added from any dissipative evolution. In practice we will focus on states with $\braket{p}_{t=0} = 0$ and small times, so this is just the variance in the momentum. We are specializing to just the axis defined by the two mass centers for simplicity. The time derivative of this is straightforward to compute using the Lindbladian evolution under Eq.~\eqref{eq:lind_restr}. One finds easily that
\begin{equation}
\label{eq:SFF_one}
\frac{d\angbracket{p_a^2}_e}{dt}=\int d^3\mb k \, k_x^2\, f_a(k).
\end{equation}
This quantity has the same units as a force noise spectral density $S_{FF}$, which indicates how it can be measured~\cite{Clerk:2008tlb,Beckey:2023shi}.\footnote{Concretely, for noise sufficiently flat in frequency (white noise), $d\braket{p^2}/dt = S_{FF}(0)$ where $S_{FF}(0)$ is the force noise spectral density at zero frequency.} It is also convenient to define a version of this with units of a rate, by normalizing with the zero point fluctuations $p_{a,0} = \sqrt{m_a \omega_a}$:
\begin{equation}
\Gamma_a = \frac{1}{p_{a,0}^2} \frac{d \braket{p_a^2}_e}{dt}.
\end{equation}
These quantities refer to single objects $a=1,2$. We also define a ``total noise'' by simply adding them:
\begin{equation}
\label{eq:SFF}
S_{FF,e} = \frac{d\angbracket{p_1^2}_e}{dt} + \frac{d\angbracket{p_2^2}_e}{dt}, \ \ \ \Gamma_{e} = \Gamma_1 + \Gamma_2.
\end{equation}
To measure these excess quantities in practice, one would have to compute the pure Hamiltonian contributions like $\braket{p^2}_H$ and then subtract them from the measured values of $\braket{p^2}$ to obtain the excess $\braket{p^2}_e$.

Now we need to quantify the entanglement. For a pair of oscillators, a convenient criterion due to Simon and based on the partial transpose will suffice~\cite{Simon:1999lfr,Kafri:2013wxa,Angeli:2025ojs}. Define a vector $\mb{M}$ consisting of the canonical variables normalized by their zero-point fluctuations:
\begin{equation}
\mb{M} = \left(\frac{x_1}{x_{1,0}},\frac{p_1}{p_{1,0}},\frac{x_2}{x_{2,0}},\frac{p_2}{p_{2,0}} \right),
\end{equation}
where $x_{a,0} = 1/\sqrt{m_a \omega_a}$. Consider the covariance matrix for an arbitrary state $\rho$:
\begin{equation}
\label{eq:covmatdef}
    \gamma_{ij}=\text{Tr}\{\rho[M_iM_j+M_jM_i]\}-2\text{Tr}\{\rho M_i\}\text{Tr}\{\rho M_j\}.
\end{equation}
Then an arbitrary (not necessarily Gaussian) state $\rho$ is entangled if
\begin{equation}
\label{eq:two_osc_ent_cond}
    \tilde{\gamma}+i\Delta_2 \ngeq 0 \, ,
\end{equation}
where these are matrices defined by
\begin{align}
\tilde{\gamma}=K\gamma K, \ \ K=\begin{pmatrix} 1 & & & \\ & 1 & & \\ & & 1 & \\ & & & -1 \end{pmatrix}, \ \ \Delta_2 = \begin{pmatrix} 0 & 1 & 0 & 0 \\ -1 & 0 & 0 & 0 \\ 0 & 0 & 0 & 1 \\ 0 & 0 & -1 & 0 \end{pmatrix}.
\end{align}
The inequality \eqref{eq:two_osc_ent_cond} is satisfied if $\tilde{\gamma}+i\Delta_2$ has at least one negative eigenvalue. 
Taking the initial condition that both oscillators are in their ground state, a short calculation shows that the lowest eigenvalue of this matrix has time derivative
\begin{equation}
\label{eq:dlambda_simple}
    \frac{d\lambda}{dt}\Big|_{t=0} = \frac{1}{2} \left( \Gamma_1+\Gamma_2-\sqrt{(\Gamma_1-\Gamma_2)^2 + 4 \Gamma_{12}^2+16g^2} \right),
\end{equation}
where $g=\alpha_G/p_{1,0}p_{2,0}$  is the gravitational coupling normalized by the zero-point fluctuations, which makes it a rate, and 
\begin{equation}
\Gamma_{12} = \frac{1}{p_{1,0} p_{2,0}} \frac{d \braket{p_1 p_2}}{dt} = \frac{\beta}{p_{1,0} p_{2,0}}
\end{equation}
represents \emph{correlated} noise acting on both masses, due to the $\beta$ term in Eq.~\eqref{eq:lind_restr}. 

With these results, we can compare our noise rates to the entanglement rates. From Eq.~\eqref{eq:dlambda_simple} we see that entanglement will be generated if
\begin{equation}
\label{eq:ent_cond_oscillators}
\Gamma_1 + \Gamma_2<\sqrt{(\Gamma_1 - \Gamma_2)^2 + 4\Gamma_{12}^2+16g^2}.
\end{equation}
The interpretation is mostly straightforward. If the single-body decoherence rates $\Gamma_a$ are too large compared to the entangling interaction strength $\alpha_G$, then no net entanglement can be generated. The role of the non-local noise term proportional to $\beta$ is more subtle: it contributes positively to the entanglement generation rate.\footnote{Note however that $\beta$ is not an independent parameter. As shown in Appendix~\ref{app:two_oscillators}, it must satisfy
$(\Gamma_1+\Gamma_2)^2\ge(\Gamma_1-\Gamma_2)^2+4\Gamma_{12}^2$, a consequence of requiring the density matrix to remain positive. As a result, the non-local $\beta$ noise term alone does not generate initial negativity. Its effect is instead to relax the threshold for entanglement generation in the presence of a genuine entangling interaction.} This is consistent with interpreting the $\beta$-term as arising from correlated environmental monitoring of the two masses, rather than from two independent local noise channels. This non-local noise term has generally been absent in other treatments of these problems.

Finally, we can use Eq.~\eqref{eq:ent_cond_oscillators} to make some numerical estimates about what noise rates would need to be measured to check if the interaction is entangling. Note that all the terms in the square root of Eq.~\eqref{eq:ent_cond_oscillators} are positive, so to make a conservative threshold we can simply drop the model-dependent noise terms. We conclude that if
\begin{align}
\label{eq:SFF_thresh}
S_{FF,e} < \frac{4G_Nm^2}{d^3} \approx 2.8 \times 10^{-47}~\frac{\textrm{N}^2}{\textrm{Hz}} \times \left(\frac{m}{1\, \textrm{mg}}\right)^2 \left(\frac{1 \, \textrm{mm}}{d}\right)^3  ,
\end{align}
then the gravitational interaction generates entanglement. Here we took $m_1 = m_2$ and $\omega_1 = \omega_2$ for simplicity. The benchmark values are modeled after state-of-the-art gravitational force measurements with oscillators~\cite{Westphal:2020okx,Agafonova:2024evr,Manley:2026evd}. The numerical value here is how we estimated the required acceleration sensitivity quoted in the introduction, since $S_{aa} = S_{FF}/m^2$.

The interpretation of Eq.~\eqref{eq:SFF_thresh} requires a bit of care. The simple intuition is that if the uncorrelated gravitationally-induced noise is larger than the gravitationally-induced entanglement rate---which is what the quantity on the right-hand side of Eq.~\eqref{eq:SFF_thresh} represents---then there is no net entanglement rate. Conversely, if the gravitationally-induced noise is smaller, then the interaction \emph{must be entangling}. 

To be more precise, note that the noise is measured in terms of single-body objects and thus independent of the distance $d$ between them. On the other hand, the entanglement rate, and thus our noise thresholds, depends on $d$. So what the bound in Eq.~\eqref{eq:SFF_thresh} really says is that if one does an experiment and rules out noise above some numerical value, then the gravitational interaction between two masses at the corresponding distance $d$ must generate entanglement.

The upshot of this is that we have two possible paths to test gravitational entanglement. One can try to look for gravitational entanglement directly, or by upper-bounding this noise. The fact that the two rates are parametrically the same suggest, however, that neither option is parametrically easier in an actual experiment. We make some further comments on this in the conclusions.

In this discussion, we used the simplified Lindbladian evolution shown in Eq.~\eqref{eq:lind_restr}. However, it is also possible to prove the same statements using the fully general form of Eq.~\eqref{eq:diss_main}, including both the cases where the Newton term arises as a Hamiltonian and as a dissipative effect. The calculations are substantially more involved, so we have relegated them to Appendix~\ref{app:two_oscillators}. Thus the entanglement inequality shown in Eq.~\eqref{eq:SFF_thresh} applies to any model of gravity obeying Galilean invariance, time locality, and the Ehrenfest conditions spelled out in Sec.~\ref{sec:model construction}.

A slightly less general version of this result was first obtained in \cite{Kafri:2013wxa} and recently revisited in \cite{Angeli:2025ojs}. Both of these works restrict the dynamics to be Gaussian, requiring the Hamiltonian to be at most quadratic in the phase-space coordinates and the Lindblad operators to be at most linear. Under this assumption, if the initial state is also Gaussian, condition \eqref{eq:two_osc_ent_cond} becomes both necessary and sufficient for entanglement generation. However, as we will demonstrate through specific examples in Section \ref{sec:examples}, the Gaussian approximation is not justified \textit{a priori} for several alternative gravity models, and the full non-Gaussian dynamics must be retained in order to set meaningful bounds on these models.

\subsection{Masses with two position states}
\label{sec:qubits}

We now turn to the case of a pair of masses which are each constrained to a pair of approximate position eigenstates. This mimics the setup proposed by Bose and collaborators~\cite{Bose:2017nin,Marletto:2017kzi} to look for gravitational entanglement generation. In their proposal, a pair of free-falling masses are both initially prepared in superpositions of the form $\ket{L}+\ket{R}$, where $\ket{L,R}$ are nearly classical trajectories centered at positions $\mb{x}_{L,R}$. By nearly classical, we mean that the wave packet sizes are much smaller than the superposition scale $\delta x = |\mb{x}_L - \mb{x}_R|$. See Fig.~\ref{fig:cartoons}, center. Again we focus on the structure of the argument and main results, and defer detailed calculations to Appendix~\ref{app:two_qubits}.

The logic is identical to the previous case. We define and compute a noise rate, an entanglement rate, and then compare them. Since we are approximating the positions as just taking two values, we can write the position operators as
\begin{equation}
x_1=-\frac{d}{2}+\frac{\delta x}{2} \sigma^z_1\, , \ \ \ x_2=\frac{d}{2}+\frac{\delta x}{2} \sigma^z_2\, ,
\end{equation}
where $\sigma^z = \ket{L}\bra{L} - \ket{R} \bra{R}$ for each mass. To characterize the noise on each mass, we can use a phase contrast. Specifically, let
\begin{equation}
\sigma_- = \frac{\sigma^x - i \sigma^y}{2} = \ket{L}\bra{R},
\end{equation}
which measures the interference contrast between the two branches of a superposition $\ket{L} + \ket{R}$. This can be measured, for example, with a standard Ramsey protocol. To characterize noise (i.e., the contribution from dissipative dynamics beyond the unitary gravitational evolution), we define the ``excess'' quantity as above:
\begin{equation}
\braket{\sigma_-}_e = \braket{\sigma_-} - \braket{\sigma_-}_H, \ \ \ \braket{\sigma_-}_H = {\rm tr} \left( \rho_H \sigma_-\right)\,,
\end{equation}
where the $H$ terms mean evolution with no Lindblad dissipation. We define the noise rate
\begin{equation}
\label{eq:noise_single_qubit}
   \Gamma_a = \left|\frac{d|\angbracket{\sigma_a^-}_e|}{dt}\right| = \int d^3\mb k \, f_a(k)\sin^2\left(\frac{k_x\delta x}{2}\right) .
\end{equation}
where the first equation is the definition and the second follows from Eq.~\eqref{eq:lind_restr}. The total noise is simply
\begin{equation}
\label{eq:gamma_noise_qubits}
\Gamma_{e} = \Gamma_1 + \Gamma_2
\end{equation}
just as in the two-oscillator case of the previous section. Again, notice that this does not involve the non-local noise coefficient $\beta$.

Now we want to compare this to the rate of entanglement generation. We will quantify this using the negativity, which for our simple two-state approximation is easy to calculate. The negativity is defined by
\begin{equation}
\label{eq:negdef}
    \mathcal{N}=\sum_i \frac{|\lambda_i|-\lambda_i}{2} \, ,
\end{equation}
where $\lambda_i$ are the eigenvalues of the partial transpose of the density matrix $\rho$. The state $\rho$ is entangled if and only if $\mathcal{N} > 0$~\cite{Peres:1996dw}. With our simple Hamiltonian, it is easy to see that the initial product state which will have the largest rate of change of the negativity is 
\begin{equation}
\label{eq:bose_init_state}
    \ket{\psi(0)}=\frac{1}{2}(\ket{L}+\ket{R})\otimes (\ket{L}+\ket{R}) \, ,
\end{equation}
which is also the presumed initial state for most experimental tests of gravitational entanglement. With this initial state and under the general time evolution of Eq.~\eqref{eq:lind_restr}, the rate of change of $\mathcal{N}$ can be computed easily:
\begin{align}
\begin{split}
    \frac{d\mathcal{N}}{dt}\Big|_{t=0} = & \frac{1}{2} \Bigg[ \sqrt{(\Gamma_1-\Gamma_2)^2+\frac{1}{4} \beta^2\delta x^4+ \alpha_G^2 \delta x^4} \\
    & - (\Gamma_1+\Gamma_2) \Bigg] \, ,
\end{split}
\end{align}
or zero if the right hand side is negative. The right-hand side is positive, that is, entanglement is generated, if
\begin{equation}
    \label{eq:qubi_cond}
    \Gamma_1+\Gamma_2 < \sqrt{(\Gamma_1-\Gamma_2)^2+\frac{1}{4} \beta^2\delta x^4+ \alpha_G^2 \delta x^4} \, .
\end{equation}
The interpretation is identical to the two-oscillator argument. If the decoherence rates $\Gamma_a$ are too large compared to the entangling interaction strength $\alpha_G$ and correlated noise $\beta$, then no net entanglement can be generated. 

Finally, we can make some conservative numerical estimates. Just like the oscillator case, all the terms in the square root in Eq.~\eqref{eq:qubi_cond} are positive. We thus see that if the excess noise satisfies $\Gamma_{e} < \alpha_G \delta x^2$, or more explicitly if
\begin{align}
\begin{split}
\label{eq:bound_qubits_simple}
\Gamma_{e} & < \frac{G_N m_1 m_2 \delta x^2}{d^3} \\
& \approx 6~{\rm Hz} \times \left( \frac{m}{10~{\rm fg}} \right)^2 \left( \frac{ \delta x}{100~{\rm \mu m}} \right)^2 \left( \frac{1~{\rm mm}}{d} \right)^3,
\end{split}
\end{align}
then the gravitational interaction is entangling. We have chosen the benchmark parameters here to match those in the Bose et al. proposal~\cite{Bose:2017nin}.

\subsection{Mechanical mass coupled to two-state system}
\label{sec:osc_qubit}

Finally, we turn to a hybrid type of experiment involving a mechanical source mass coupled to a two-state system. See Fig.~\ref{fig:cartoons}, right. This kind of setup was proposed explicitly as a gravitational noise measurement in~\cite{Carney:2021yfw}, where the source is a torsion pendulum and the two-state system is an lattice atom interferometer, where the position of the atoms can again be described by a two-state system. Other atom interferometer experiments with either lab-based sources~\cite{asenbaum2017phase,overstreet2022observation} or simply the field of the Earth~\cite{kovachy2015quantum} can similarly be described this way.

The logic follows identically to the previous two cases, so we simply quote the basic result here. Detailed computations can be found in Appendix~\ref{app:hybrid}. The noise observable is simply the sum of the two kinds of quantities from the previous sections:
\begin{equation}
\Gamma_e = \left|\frac{d \braket{|\sigma_-|}_e}{dt}\right| + \frac{1}{p_0^2} \frac{d \braket{p^2}_e}{dt}.
\end{equation}
Again, this can be measured independently on each device and does not require correlated readout. In turn, the partial transpose (on the two-state system) can again be used to determine if the joint state is entangled: if the partial transposed density matrix has a negative eigenvalue, then the joint state is entangled~\cite{Peres:1996dw}. 

Assuming that the initial state of the system is of the form
\begin{equation}
    \ket{\psi(0)}=\frac{1}{\sqrt{2}}(\ket{L}+\ket{R})\otimes \ket{0} \, ,
\end{equation}
where $\ket{0}$ is the ground state of the pendulum, then if
\begin{align}
\begin{split}
   \Gamma_e & < \frac{2G_NmM\delta x}{d^3\sqrt{2M\omega}}  \\
   & \approx 10^{-16}~{\rm Hz} \times \left(\frac{\delta x}{1~{\rm \mu m}}\right)\left(\frac{1~\textrm{mm}}{d}\right)^3\left(\frac{1~\textrm{Hz}}{\omega}\right)^{1/2}.
\end{split}
\end{align}
then the interaction must be entangling. Here we have taken $M = 1~{\rm g}$ and $m = m_{\rm Cs} = 132~{\rm u}$ for the mass benchmarks. The small atomic mass is the reason for the small numerical value of the noise bound.

\section{Bounds on specific gravity models}
\label{sec:examples}

We now turn to the analysis of some specific models of gravity that have been proposed as alternatives to perturbatively quantized gravity. These models each have some freedom---free parameters and/or functions---which can be constrained by experiments. They also have a variety of entanglement generation rates. One of the key goals of this section is to show that there is a finite range of currently unexcluded models that can be constrained by noise searches. This can be done continuously as noise searches become more sensitive, meaning that useful information can be gained even before experiments reach the entanglement noise thresholds quoted in the previous section.

\subsection{``Classical-quantum'' gravity}
First we consider a recently proposed model due to Oppenheim and collaborators, which they refer to as ``classical-quantum'' (CQ) gravity~\cite{Oppenheim:2018igd,Layton:2023oud,Oppenheim:2023izn}. In this model, the gravitational field is taken to be a fundamentally classical, albeit stochastic, system. Coupling this field to quantum matter inevitably introduces decoherence on the matter itself. 

One of the most important properties of the CQ framework is that it can be defined relativistically, using a path integral. For our purposes, however, we can work in the non-relativistic limit, which is described by a Lindblad evolution for massive objects once the gravitational field is averaged out. We will present this in the simple form shown in~\cite{Carney:2024izr}. The masses obey the evolution law
\begin{equation}
\label{eq:Opp_Lind_main}
\begin{aligned}
    &\dot \rho=-i[H,\rho]\\
    & + \int d^3\mb x\, d^3 \mb y \, F_2(\mb x-\mb y)\left[J(\mb x)\rho J(\mb y)-\frac{1}{2}\{J(\mb x)J(\mb y),\rho \}\right] \, ,
    \end{aligned}
\end{equation}
where $H$ is the usual Hamiltonian for masses coupled by the Newtonian potential [Eq.~\eqref{eq:ham_initial}], and the dissipation terms are defined by
\begin{align}
\begin{split}
   F_2(\mb x)& = D_2 G_C(\mb x)+D_0\delta^3(\mb x) \\
   G_C(\mb x) & =\int \frac{d^3\mb k}{(2\pi)^3}\frac{e^{i\mb k\cdot \mb x}}{(\mb k^2+m_\phi^2)^2} \\
    J(\mb x) & =\sum_{a} \lambda_a \delta^3(\mb x-\mb x_a).
\end{split}
\end{align}
Here, $\lambda_a=\sqrt{4\pi G_N}m_a$ are the coupling constants, and $m_\phi \to 0$ is an infrared regulator.\footnote{The notation $\lambda$ and $\phi$ here reflect how this master equation was derived in~\cite{Carney:2024izr}, which is by replacing the full gravitational field with a simple Yukawa-type mediator. In the non-relativistic limit these give the same results.} The $\mb{x}_a$ are the position operators, while the rest of the $\mb{x},\mb{y}$ are just coordinates. 
$D_0,D_2 > 0$ are the fundamental free parameters of CQ gravity, and are called the decoherence and diffusion parameters, respectively. As shown in~\cite{Oppenheim:2018igd,Layton:2023oud,Oppenheim:2023izn}, they must satisfy a specific relation (``decoherence-diffusion tradeoff'')
\begin{equation}
\label{eq:diff_dec_tradeoff}
    D_0D_2\geq 1
\end{equation}
for the model to give well-defined evolution. Thus in total the CQ model, as presented here, has a two-dimensional parameter space.

The form of Eq.~\eqref{eq:Opp_Lind_main} is clearly Galilean invariant, but otherwise does not manifestly match our general framework from Sec.~\ref{sec:model construction}. A few short calculations (see Appendix~\ref{app:cq_gravity}) show that we can reduce the Lindblad equation to the simple form of Eq.~\eqref{eq:lind_restr}, with
\begin{align}
\begin{split}
\label{eq:CQ_factors}
f_a(k) & = \frac{G_N m_a^2}{2\pi^2}\left[\frac{D_2}{(\mb k^2+m^2_{\phi})^2}+D_0\right]  \\
\beta & = -\frac{2G_Nm_1m_2(D_0+D_2\ell^4)\cos(d/\ell)}{\pi\ell^3 d^2}
\end{split}
\end{align}
Thus our general results from the previous sections can be applied. In this expression, $\ell$ is a small length scale that we have to introduce as an ultraviolet regulator. When applying these results to composite objects (for example, masses in a lab), $\ell$ can be taken to be of order the physical size of the objects.

\begin{figure}
    \centering
    \includegraphics[width=1\linewidth]{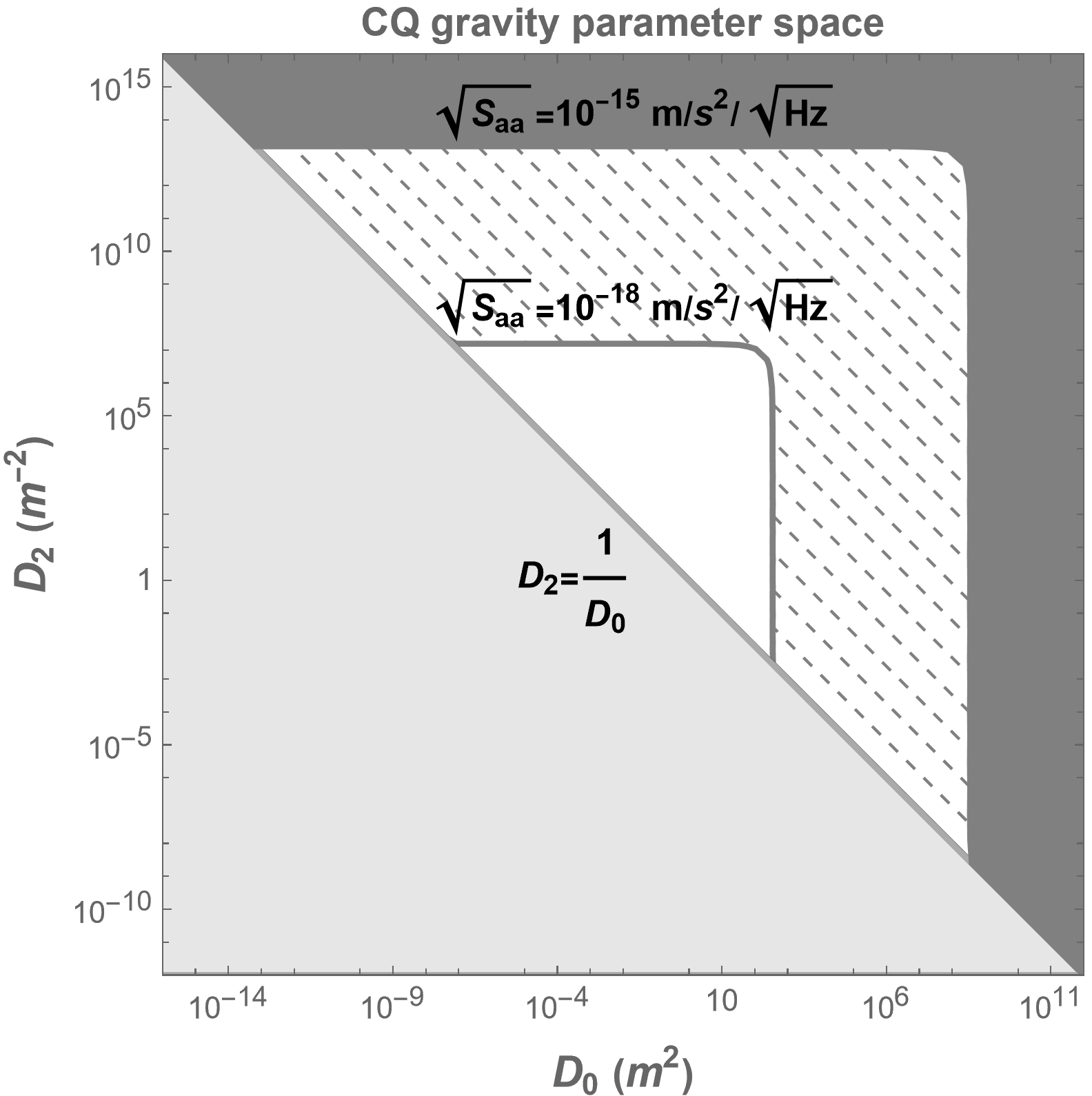}
    \caption{\textbf{The parameter space of classical-quantum gravity.} Here we show the two-dimensional space of free parameters $D_0$, $D_2$ that define CQ gravity.  The lower left region in gray violates the basic tradeoff relation $D_0 D_2 \geq 1$ and is thus not part of the parameter space. The top right dark gray region is already excluded because it would produce more noise [see Eq.~\eqref{eq:cq_forcenoise}] than observed in the LISA Pathfinder (a device of linear size $\ell \approx 46~{\rm mm}$), which operated with acceleration noise $\sqrt{S_{aa}} \lesssim 10^{-15}~{\rm m/s^2}/\sqrt{\rm Hz}$~\cite{armano2024depth}. The dashed region shows the part of this parameter space would be ruled out by comparing the force noise predicted by Eq.~\eqref{eq:cq_forcenoise} to a future acceleration measurement at the general non-entangling threshold $S_{aa} \approx 10^{-18}~{\rm m/s^2}/\sqrt{\rm Hz}$ [see Eq.~\eqref{eq:SFF_thresh}]. To render this plot, we have not assumed that CQ is non-entangling; if we had, then the measurement at the $10^{-18}$ level would be sufficient to completely rule out the model. The bounds obtained are sensitive to the choice of the cutoff scale $\ell$, which in this example we have taken to be $\ell\approx 46 \,\textrm{mm}$, in connection with current and possible future LISA measurements.}
    \label{fig:cq}
\end{figure}

To begin, we can use Eq.~\eqref{eq:CQ_factors} to compute a typical noise level in CQ gravity. Since mechanical oscillators are currently the most accurate force sensors available, we can compute the force noise $S_{FF}$ in CQ theory. Plugging Eq.~\eqref{eq:CQ_factors} into Eq.~\eqref{eq:SFF}, we obtain
\begin{align}
\begin{split}
    \label{eq:cq_forcenoise}
    S_{FF,e} & =\frac{4G_N m^2}{15\pi \ell^5}\left[ D_0+5 \ell^4 D_2\right] \\
    & \approx \frac{4 G_N \rho_{\rm solid}^2}{15 \pi} \left[ \ell D_0 + 5 \ell^5 D_2 \right].
\end{split}
\end{align}
This depends sensitively on our cutoff scale $\ell$. At present, we do not know how to renormalize CQ theory appropriately (see~\cite{Grudka:2024llq} for one attempt), so our estimates in what follows should be considered somewhat phenomenological. To make a concrete estimate in the second line, we took $\ell$ to be of the size of the physical objects under consideration, so that everything can be rewritten in terms of the solid density $\rho_{\rm solid} = m/\ell^3$. With this choice, $S_{FF} \to 0$ as $\ell \to 0$. We show some numerical estimates in Fig.~\ref{fig:cq} with typical state-of-the-art force sensors.

Naively, the CQ model should not be able to entangle pairs of objects, since they only interact through a classical medium~\cite{Oppenheim:2018igd,Layton:2023oud,Oppenheim:2023izn}. We can see a remnant of this idea in our simple framework here. The diffusion-decoherence tradeoff [Eq.~\eqref{eq:diff_dec_tradeoff}] implies that the force noise in Eq.~\eqref{eq:cq_forcenoise} cannot be arbitrarily small. To make this more precise, we have $ D_0 +5 \ell^4 D_2 \geq 2\sqrt{5\ell^4 D_0 D_1} \geq 2\sqrt{5}\ell^2$ using the arithmetic–geometric mean inequality and Eq.~\eqref{eq:diff_dec_tradeoff}. Inserting this into Eq.~\eqref{eq:cq_forcenoise} we find that CQ gravity has force noise
\begin{align}
\label{eq:cq_mininoise}
    S_{FF,e} \geq  \frac{8 \sqrt{5} G_N m^2}{15 \pi \ell^3} \gtrsim \frac{4G_N m^2}{d^3}.
\end{align}
Since $\ell$ denotes the size of a mass, any experiment with two masses separated by $d$ should obviously satisfy $d \gtrsim {\rm few} \times \ell$, which is what leads to the final inequality. In other words, CQ gravity always produces noise above the entangling threshold of Eq.~\eqref{eq:SFF_thresh}.

There are thus three somewhat different paths to ruling out CQ gravity. One would be to simply observe noise below the value predicted by Eq.~\eqref{eq:cq_forcenoise}. In Fig.~\ref{fig:cq}, we show the parameter space excluded in this way by current low-noise accelerometer measurements. The other would be to take for granted that CQ gravity cannot entangle masses. This is a pure theory question about the model; we believe it is true, but there has been some recent controversy around the issue \cite{Aziz:2025ypo,Diosi:2025hhe,Marletto:2025fpm}. Assuming that CQ gravity cannot entangle would imply that $S_{FF,e} > S_{FF,{\rm thresh}}$, so then a measurement with noise below $S_{FF,{\rm thresh}}$ would also rule out CQ gravity. Finally, of course, we could also assume CQ gravity cannot entangle masses, and then just directly measure entanglement generation.

We end our discussion of the CQ model with a technical methodological comment. One of the motivations for this paper, and specifically our general evolution framework described in Sec.~\ref{sec:model construction}, was that some basic effects in the CQ model cannot be accurately captured by the Gaussian evolution laws used in~\cite{Kafri:2013wxa,Oppenheim:2022xjr,Lami:2023gmz,Angeli:2025ojs}. Concretely, consider a \emph{single} isolated mass $m$ with position $\mb{x}$. If we place this mass in superposition of two locations separated by distance $\delta x$, say of the form $\ket{L} + \ket{R}$, the noise in CQ gravity will cause this superposition to decohere. The decoherence rate can be calculated using Eqs.~\eqref{eq:noise_single_qubit} and \eqref{eq:CQ_factors}, which gives
   \begin{equation}
   \Gamma \approx \frac{G_N m^2}{2\pi^2} \left[ \frac{D_0}{\ell^3} + 4\pi^2 D_2 \delta x \right].
\end{equation}
We have not been careful with $\mathcal{O}(1)$ factors here. The point we want to emphasize is that if we had approximated the Lindblad equation as quadratic in the $\mb{x}$ operators, it would predict $\Gamma \sim \delta x^2$, which is parametrically the wrong answer.

\subsection{Entropic gravity}
\label{sec:entropic}

Now we turn to some phenomenology based on the notion that gravity could arise as an effective thermodynamic or ``entropic'' force. This idea was first postulated quite generally by Jacobson~\cite{Jacobson:1995ab}, with the specific idea of an entropic force later suggested by Verlinde~\cite{Verlinde:2010hp}. 

Recently, some specific microscopic realizations of this idea were proposed in~\cite{Carney:2025wmz}, and in this section we analyze their phenomenology. These entropic models are only in the Newtonian limit and should be viewed as toy models or examples of the broader idea of gravity as an entropic interaction. However, they are explicit enough that they give concrete predictions for the noise generated by the thermal fluctuations in the gravitational interaction, and this is what we study in this section. They also give a nice counter example to the notion that gravity is ``either classical or quantum''. The model is a completely standard quantum many-body system, but at long distances looks like a semiclassical gravitational interaction plus small fluctuations which are distinct from gravitons. \\

\textbf{Non-local model.} We first analyze the ``non-local'' model of~\cite{Carney:2025wmz} because it is simple and highlights a number of properties of the general framework presented in Sec.~\ref{sec:model construction}. The non-local model realizes Newtonian two-body interactions by directly coupling a set of masses with positions $\mb{x}_a$ to a thermal mediator system directly through their relative positions $\mb{x}_{ab} = \mb{x}_a - \mb{x}_b$. In the concrete realization in~\cite{Carney:2025wmz} this is achieved by using a large collection of qubits for the mediators and having the qubit frequencies $\omega_{\rm med} = \omega_{\rm med}(\mb{x}_{ab})$ depend on the relative mass positions. This is non-local in the same sense as the ordinary Newtonian potential $V_N \sim 1/|\mb{x}_{ab}|$.

With this non-local type of coupling, the resulting evolution equations can be analyzed easily and in a fully Gaussian approximation. The general evolution for the masses takes the form
\begin{equation}
\label{eq:lindblad_entropic}
    \dot{\rho} =-i[H,\rho] +\sum_{\alpha,\pm}\left[K_{\alpha,\pm}\rho K_{\alpha,\pm}^\dagger-\frac{1}{2}\{K_{\alpha,\pm}^\dagger K_{\alpha,\pm},\rho\}\right]
\end{equation}
where once again $H$ contains the single-mass energies and the Newtonian potential given in Eq.~\eqref{eq:ham_initial}. The noise operators $K_{\alpha,\pm} = K_{\alpha,\pm}(\mb{x}_{ab})$ represent thermal fluctuations of the gravitational interaction. For an experiment involving two masses whose fluctuations are small compared to their baseline separation $d$, we can simply Taylor expand the relative displacement. We write $\mb{r} = \mb{d} + \mb{x}_1 - \mb{x}_2$ and expand the Lindblad operators
\begin{equation}
K_{\alpha,\pm}(\mb{r}) \approx K_{\alpha,\pm}(\mb{d}) + \nabla K_{\alpha,\pm}(\mb{d}) \cdot \Delta \mb{x}
\end{equation}
with $\Delta \mb{x} = \mb{x}_1 - \mb{x}_2$. The Lindblad equation then takes the very simple form
\begin{align}
\label{eq:entropic_nonlocal_lindblad}
\dot{\rho} = - i[H_0,\rho] - i \alpha_G [\Delta x^2,\rho]  + \beta \left( \Delta x \rho \Delta x - \frac{1}{2} \left\{ \Delta x^2, \rho \right\} \right),
\end{align}
where
\begin{equation}
\beta = \sum_{\alpha,\pm} \left( \partial_x K_{\alpha,\pm}(\mb{d}) \right) \left( \partial_x K_{\alpha,\pm}(\mb{d}) \right)
\end{equation}
and again we are restricting to the single spatial axis $x$ along $\mb{d}$. This is in the same form as Eq.~\eqref{eq:lind_restr}, with the extra simplification that the single-body dissipation terms involving $f_a(k)$ functions are reduced to simple constants with the same strength as the correlated two-body noise term $\beta [x_1,[\rho,x_2]]$. Moreover, the entire evolution is Gaussian.

Evolution of the masses under the non-local entropic model is very interesting. It violates some common lore about models where gravity is not quantized into gravitons. Specifically, consider the kind of experiment involving two masses in two-state superpositions with separation scale $\delta x$, as described in Sec. \ref{sec:qubits} and shown in Fig.~\ref{fig:cartoons}, center. The dissipation terms in Eq. \eqref{eq:entropic_nonlocal_lindblad} give single-mass noise rates, using Eq.~\eqref{eq:noise_single_qubit},
\begin{equation}
    \Gamma_1=\Gamma_2=\frac{\beta \delta x^2}{4} \, .
\end{equation}
So, in particular, the total excess noise $\Gamma_e = \Gamma_1 + \Gamma_2 = \beta \delta x^2/2$, so obviously
\begin{equation}
    \Gamma_{1} + \Gamma_{2} <\sqrt{\frac{\beta^2 \delta x^4}{4}+\alpha_G^2\delta x^4}.
\end{equation}
But this is even more stringent than the entanglement threshold in Eq.~\eqref{eq:qubi_cond}. Thus we conclude that the non-local entropic gravity model \emph{does} produce entanglement between massive objects, independent of any free parameter choices in the model.\footnote{We emphasize that this conclusion is not tied to the specific setup of the entropic gravity non-local model. Any non-quantum gravity model with noise of the form \eqref{eq:entropic_nonlocal_lindblad} is guaranteed to generate entanglement during the evolution.} 

Thus, experiments purely looking at direct entanglement generation will not be able to exclude such a model in favor of perturbative quantum gravity. However, it should be emphasized that the actual pattern of entanglement differs, so a specific choice of entanglement witness could distinguish the two cases. For example, consider the entanglement witness operator proposed in~\cite{Bose:2017nin}:
\begin{equation}
    W=X_1\otimes Z_2+Y_1\otimes Y_2 \, ,  
\end{equation}
for use in a measurement in the two-state approximation considered in Sec.~\ref{sec:qubits}. If $|\angbracket{W}|>1$, then the state in which $W$ has been measured must be entangled. Perturbative quantum gravity predicts that $|\angbracket{W}| > 1$ will occur for the initial condition shown in Fig.~\ref{fig:cartoons}, i.e., with both masses prepared in two-state superpositions. The entropic model also generates entanglement, but the detailed prediction for $|\angbracket{W}|$ with the same initial condition is~\cite{Carney:2025wmz}
\begin{equation}
\label{eq:witness_non_local}
\begin{aligned}
    |\angbracket{W}|=\frac{1}{2}\Big|&e^{-4\Gamma \Delta t}\cos(\phi_{LR}-\phi_{RL})+\\
    &e^{-\Gamma\Delta t}(\cos\phi_{RL}-\cos\phi_{LR})-1\Big| \, .
    \end{aligned}
\end{equation}
Here $\phi_{LR}=-\alpha_G(d-\delta x)\delta x\Delta t$,  $\phi_{RL}=\alpha_G(d+\delta x)\delta x\Delta t $, and $\Gamma\sim \alpha_G\delta x^2 /(1+\lambda d/\ell^2)$ is the decoherence rate from the entropic fluctuations~\cite{Carney:2025wmz}, where $\lambda$ and $\ell^2$ are free parameters of the entropic model. For generic choices of these parameters, the value of $|\angbracket{W}|$ may be below $1$ for the entire duration of the experiment, thus prohibiting the detection of entanglement generation through this witness. A different $W$ operator would have to be chosen to detect the entanglement.

Nevertheless, a distinct prediction of models described by \eqref{eq:entropic_nonlocal_lindblad} is the force noise generated between two objects interacting gravitationally. The evolution of this excess noise can be calculated by the general methods above as we review briefly in Appendix~\ref{appendix-entropic}, or by the methods used in~\cite{Carney:2025wmz}. In any case, the result is
\begin{equation}
    S_{FF,e}=\frac{d\braket{p^2}_e}{dt}=\beta \approx \frac{G_N m_1 m_2}{d^3(1 + \lambda d/\ell^2)}.
\end{equation}
Thus measurements of force noise can be used to constrain these parameters, and in particular can be used to discriminate such a model from perturbative quantum gravity. \\

\textbf{Local model.} Finally, we briefly repeat our analysis in the ``local'' entropic gravity model of~\cite{Carney:2025wmz}. In the context of the framework in this paper, the local entropic model is qualitatively more similar to the CQ model than the non-local entropic model. It has local noise which acts to degrade entanglement generation.

In the local entropic model, the Newtonian interaction arises by assuming the existence of some space-filling medium which couples locally to the positions of the masses $\mb{x}_a$. In the concrete example in~\cite{Carney:2025wmz}, the medium was taken to be a space-filling lattice with lattice spacing $a$ and one qubit per site, with the qubit frequencies $\omega = \omega(\mb{x}_1,\mb{x}_2,\ldots)$ now depending separately on the different mass positions rather than their differences. 

The resulting Lindblad evolution for massive objects still takes the form of Eq.~\eqref{eq:lindblad_entropic}. However, the noise operators $K_{\alpha,\pm} = K_{\alpha,\pm}(\mb{x}_1,\mb{x}_2,\ldots)$ are, like the qubit frequencies, functions of the individual mass positions rather than differences. Thus in general there is no way to do a Gaussian expansion. In particular, even with just a single mass with position operator $\mb{x}$, the noise operators $K_{\alpha,\pm}(\mb{x})$ have no particular scale to compare this to, so the full non-Gaussian structure of the $f_a(k)$ functions in Eq.~\eqref{eq:lind_restr} is required to accurately describe the model. Concretely, for two masses $m_1, m_2$ separated by distance $d$ much larger than the lattice spacing $d \gg a$, one finds
\begin{equation}
\begin{aligned}
\label{eq:fbeta_local}
    &f_a(k)=\frac{\pi (\lambda_+^2+\lambda_-^2)m^2_aL^4}{2a^3}\frac{e^{-2ak}}{k^2}\\
    &\beta \approx -\frac{2\pi^3}{a^3d^3}(\lambda_+^2+\lambda_-^2)m_1m_2L^4 \, , 
    \end{aligned}
\end{equation}
where in the last line we have assumed that $d\gg a$. Here,
\begin{equation}
    \lambda_+^2=\frac{\sigma_*\gamma_{th}}{4T^2}\,, \qquad \lambda_-^2=\frac{2\sigma_*(\sigma_*-1)^2}{\gamma_{th}}
\end{equation}
and $T$, $\sigma_*$, $\gamma_{\rm th}$, and $L$ are free parameters of the model, describing the mediator qubits' temperature, average polarization, thermalization rate, and an overall length scale, respectively.

The level of noise introduced by the thermal fluctuations in this model can now be computed easily. For definiteness, consider a pair of oscillators of mass $m$ separated by an equilibrium distance $d$. Using Eq.~\eqref{eq:fbeta_local} in Eq.~\eqref{eq:SFF_one}, we find
\begin{equation}
\label{eq:SFF_local_entropic}
    S_{FF,e} =\frac{G_N m^2}{12 \pi a^3}\left(\frac{1}{\eta}+8\eta\right) \geq \frac{G_N m^2}{a^3} \frac{ \sqrt{2}}{3\pi},
\end{equation}
where $\eta=T(1-\sigma_*)/\gamma_{\rm th}$ and the second result is a simple minimization over this factor. This increases quickly with smaller lattice spacing $a \to 0$, which allows us to set a stringent lower bound on $a$ by comparing to low-noise force sensing experiments. In~\cite{Carney:2025wmz}, the bound $a \gtrsim 10^{-13}~{\rm m}$ was given using roughly similar logic applied to anomalous heating. Using the specific logic shown here, we can substantially improve this bound: converting this to an acceleration noise, we have (with apologies for the double use of the symbol $a$)
\begin{equation}
   \sqrt{ S_{aa}} = \sqrt{\frac{S_{FF,e}}{m^2}} \gtrsim 1 \times 10^{-15}~\frac{\rm m/s^2}{\sqrt{\rm Hz}} \times \left( \frac{10~{\rm \mu m}}{a} \right)^{3/2}.
\end{equation}
Here we benchmarked the noise to the LISA pathfinder acceleration sensitivity~\cite{armano2024depth}. Thus we require the lattice to be no smaller than about $a \sim 10~{\rm \mu m}$, otherwise LISA Pathfinder would have seen more noise than it did. In~\cite{Carney:2025wmz} it was also claimed that lattice spacings $a \gtrsim 100~{\rm \mu m}$ would lead to observable deviations from a strict $1/r$ falloff of the gravitational field in torsion balance experiments~\cite{Adelberger:2003zx,lee2020new}, so this leaves a very small healthy parameter space $10~{\rm \mu m} \lesssim a \lesssim 100~{\rm \mu m}$.

Finally, we can compare these noise levels to our entanglement thresholds. The naive expectation would be that a highly local, thermodynamic interaction cannot entangle massive systems, and indeed one can easily see that this is true in this model using the results in this paper. As long as we are considering two masses separated by more than the lattice spacing $d \gtrsim a$, then Eq.~\eqref{eq:SFF_local_entropic} says that $S_{FF} \gtrsim G_N m^2/d^3$, which is precisely the entanglement threshold [see Eq.~\eqref{eq:SFF_thresh}]. Thus the local entropic model cannot produce any entanglement, and would be ruled out if gravitationally-mediated entanglement was observed.

\section{Outlook}

We are in an era in which real experiments that probe the quantum nature of gravity are becoming possible. To make these experiments as useful as possible, it is crucial to clearly map out the kinds of gravitational models and phenomenologies that they can test, and to provide sharp and quantitative experimental targets. 

In this paper, we have provided a general parametrization of the space of models of gravity, both perturbatively quantized and otherwise, applicable to experiments in a non-relativistic setting. This parametrization builds on pioneering earlier work~\cite{Kafri:2013wxa,Oppenheim:2022xjr,Lami:2023gmz,Angeli:2025ojs}, but extends it by including non-Gaussian and correlated noise effects, which turn out to be crucial to accurately capture a number of explicit gravitational models~\cite{Oppenheim:2018igd,Layton:2023oud,Oppenheim:2023izn,Kryhin:2023cyw,Carney:2024izr,Carney:2025wmz}. 

We have seen that in models which are not quantized in the usual sense of treating small gravitational field perturbations as gravitons, there is some amount of irreversibility in the time evolution under the gravitational interaction. This was expected from earlier results~\cite{Kafri:2013wxa,Oppenheim:2022xjr,Lami:2023gmz,Angeli:2025ojs} as well as on more general theoretical grounds~\cite{Carney:2021vvt,Galley:2023byb}. This irreversibility can be measured as noise, leading to heating and decoherence in experiments beyond that expected from the graviton picture. In particular, there is a certain set of noise thresholds such that if an experiment can be done to rule out gravitationally induced noise at the threshold levels, one can conclude that the gravitational interaction must entangle.

We have also added a twist to some common lore on the possible behavior of ``non-quantized'' gravitational models. Throughout this paper, by ``quantized'' gravity, we have specifically meant the usual effective quantum field theory of gravitons. It is often assumed that detecting gravitational entanglement would rule out any other possibility. In Sec.~\ref{sec:entropic}, however, we saw that the non-local entropic model also predicts the generation of entanglement, for all values of its free parameters. It is nonetheless distinct from the graviton models, because it also produces irreversibility and noise. Thus an entanglement measurement cannot distinguish the entropic and graviton models, while a sufficiently sensitive noise measurement can.

As usual, this leaves the central future direction: actual performance of such measurements. Currently, a great deal of effort is focused on performing an experiment to determine if the Newtonian gravitational interaction can entangle a pair of masses. In practice, this is an extremely challenging task. Here, we are proposing that anomalous noise searches give a complementary approach that can reach, in principle, the same conclusions. 

The obvious question is whether such a noise search is easier than a direct entanglement measurement. The naive answer is no: the noise thresholds essentially say that one has to measure noise at a level below the gravitational coupling strength, like the acceleration noise $\sqrt{S_{aa}} \lesssim \sqrt{G_N/d^3} \sim 10^{-18}~{\rm m/s^2}/{\sqrt{\rm Hz}}$ level quoted above (with $d \sim 1~{\rm mm}$, to match typical entanglement tests with mechanical systems). Less prosaically, actually building sensors with these levels of sensitivity will require substantial quantum noise engineering. This task is not parametrically easier than the direct entanglement measurements, which require a similar level of quantum control of macroscopic systems. 

Nevertheless, the quest for experimental data probing the quantum nature of gravity is both critically important and fraught with difficulty. Having multiple possible technical pathways can only be beneficial. Moreover, as experiments improve in noise sensitivity from current experiments to future ones a few orders of magnitude away at the entanglement thresholds, the kinds of bounds shown here can be used to systematically search over or exclude the remaining parameter space of the non-graviton models. We look forward to the next generation of measurements, which will take us closer to a definite statement about whether the gravitational field has the same quantum properties as the other fundamental fields in nature, or whether it behaves in some more mysterious fashion.

\section*{Acknowledgments}
We thank Lorenzo Braccini, Kai Isaak-Ellers, Manthos Karydas, Jonathan Oppenheim, Emanuele Panella, Nicolò Piccione, Roshni Singh, and Jacob Taylor for helpful discussions. Work at Berkeley Lab is supported by the U.S. Department of Energy (DOE), Office of High Energy Physics, under Contract No. DEAC02-05CH11231, by DOE Quantum Information Science Enabled Discovery (QuantISED) for High Energy Physics, and by the Heising-Simons Foundation under grant 2023-4467 ``Testing the Quantum Coherence of Gravity''. G.F. also acknowledges financial support from the Blanceflor Foundation.
T.F. and A.M. are supported by JSPS KAKENHI 
Grants Nos. JP23K03424 (T.F.), JP23K13103 (A.M.), and JP23H01175 (A.M.).

\bibliographystyle{utphys}
\bibliography{Refs}

\clearpage
\newpage 

\begin{appendix}

\onecolumngrid
\section{Detailed derivations of the Lindblad equations}
\label{app:symmetryapp}
\subsection{Constraints on the Lindblad operators}
We impose the conditions outlined in Sec. \ref{sec:model construction} to constrain the form of our general quantum channel
\begin{equation}
\label{eq:lind_appendix}
        \dt\rho=\mathcal{L}(\rho)=-i[H,\rho]+\mathcal{D}(\rho) \, ,
\end{equation}
where
\begin{equation}
\label{eq:hamdiss_app}
\begin{aligned}
    \mathcal{D}(\rho)=&\int d\lambda\left[ \, L_\lambda(\mb x_1,\mb x_2)\rho L_\lambda^\dagger(\mb x_1,\mb x_2)-\frac{1}{2}\{L^\dagger_\lambda(\mb x_1,\mb x_2)L_\lambda(\mb x_1,\mb x_2),\rho\}\right] \,,
    \end{aligned}
\end{equation}
and the Hamiltonian takes the form \eqref{eq:ham_initial} or \eqref{eq:ham_initial_2}, whether the Newtonian interaction is sourced by the Newton potential or by the dissipation terms, respectively.
Here we assume that the Lindblad operators $\hat L_\lambda$ are functions only of the position operators to connect our framework to the majority of examples models in the literature.
We start by imposing Galilean symmetry. In the context of Lindblad dynamics, we say that a unitary operator $U$ is a symmetry of the system if it maps solutions of the dynamics $\rho$ to solutions of the same dynamics, namely:
\begin{equation}
\label{eq:symmetry_cond_app}
    \dot \rho=\mathcal{L}(\rho) \Leftrightarrow \dot \rho' =\mathcal{L}(\rho') \, ,
\end{equation}
with $\rho'=U\rho U^\dagger$. This condition implies
that the Hamiltonian is invariant while the dissipative part transforms covariantly
\begin{equation}
\label{eq:hamdiss_invariance}
    H=UHU^\dagger +i(\partial_tU)U^\dagger\, , \qquad  U\mathcal{D}(\rho)U^\dagger = \mathcal{D}(U\rho U^\dagger) \, .
\end{equation}
Of course, the Hamiltonian is already invariant under Galilean symmetry and requires no further simplification. As for the dissipation term, we conveniently switch to center of mass and relative position coordinates:
\begin{equation}
    \begin{aligned}
    \mb R=&\frac{m_1\mb x_1+m_2\mb x_2}{m_1+m_2}, \qquad \mb P=\mb p_1+\mb p_2, \\
    \mb r=& \mb x_1-\mb x_2, \, \qquad  \mb p=\frac{m_2\mb p_1-m_1\mb p_2}{m_1+m_2}.
    \end{aligned}
\end{equation}
We now impose Galilean symmetry in stages, considering translations, boosts, and rotations in turn. We begin with spatial translations $U_\mb a=e^{i\mb P\cdot \mb a}$. Under this transformation, the symmetry condition \eqref{eq:hamdiss_invariance} is equivalent to
\begin{equation}
\label{eq:rep_condition}
    U_{\mb a} L_\lambda(\mb R,\mb r)U_{\mb a}^\dagger =\int d \mb \lambda' \, \mathcal{U}_{\lambda\lambda'}(\mb a)L_{\lambda'}(\mb R,\mb r) \, ,
\end{equation}
for $\mathcal{U}_{\lambda\lambda'}(\mb a)$ unitary, in the sense that 
\begin{equation}
    \int d \lambda''\,  \, \mathcal{U}_{\lambda\lambda''}(\mb a)\mathcal{U}_{\lambda'\lambda''}(\mb a)=\delta(\lambda-\lambda') \, .
\end{equation}
It can also be checked that $\int d \mb \lambda'' \, \mathcal U_{\lambda\lambda''}(\mb a)\mathcal U_{\lambda''\lambda'}(\mb a')=\mathcal U_{\lambda\lambda'}(\mb a+\mb a')$. More formally, this last condition and \eqref{eq:rep_condition} imply that $L_\lambda(\mb R,\mb r)$ is a representation of the translation group. We can write $\mathcal{U}_{\lambda\lambda'}(\mb a)$ in a diagonal form, namely
\begin{equation}
\label{eq:Udiag}
    \mathcal{U}_{\lambda\lambda'}(\mb a)=\sum_c\int d^3\mb k \, e^{i\mb k\cdot \mb a} \, u_{\lambda,c}(\mb k)u^*_{\lambda',c}(\mb k) \, , 
\end{equation}
where $c$ is a degeneracy index. The basis vectors $u_{\lambda,c}(\mb k)$ satisfy
\begin{equation}
    \sum_{c}\int d^3 \mb k \, u_{\lambda,c}(\mb k)u_{\lambda',c}^*(\mb k) = \delta(\lambda-\lambda')\, , \qquad \int d^3\lambda \, u_{\lambda,c}(\mb k)u^*_{\lambda,c'}(\mb k') =\delta^3(\mb k-\mb k')\delta_{cc'} \, .
\end{equation}
Using these completeness relations and the diagonal form of $\mathcal{U}_{\lambda\lambda'}$ in \eqref{eq:Udiag}, 
we introduce a new basis of Lindblad operators
\begin{equation}
\label{eq:V_fourier_def}
    L_{\mb k,c}(\mb R,\mb r)=\int d\lambda \, u^*_{\lambda,c}(\mb k) L_{\lambda}(\mb R,\mb r) \,.
\end{equation}
This basis is chosen so that the action of spatial translations is diagonal, and condition \eqref{eq:rep_condition} reduces to
\begin{equation}
    U_{\mb a}L_{\mb k,c}(\mb R,\mb r)U_{\mb a}^\dagger =L_{\mb k,c}(\mb R+\mb a,\mb r)=e^{i\mb k\cdot \mb a}L_{\mb k,c}(\mb R,\mb r) \, .
\end{equation}
This implies that $L_{\mb k,c}(\mb R,\mb r)$ is of the form 
\begin{equation}
\label{eq:J_def}
    L_{\mb k,c}(\mb R,\mb r)=e^{i\mb k\cdot \mb R}J_{\mb k,c}(\mb r) \,,
\end{equation}
for some unknown function of the relative position $J_{\mb k,c}(\mb r)$. 
Imposing boost symmetry, described by the unitary $U_{\mb v}=e^{i\mb v\cdot ((m_1+m_2)\mb R-\mb Pt)}$, leads to a similar analysis and does not further constrain the dissipation terms. The dissipation superoperator can then be written as
\begin{equation}
    \mathcal{D}(\rho)=\sum_i\int d^3\mb k \left[e^{i\mb k\cdot \mb R}J_{\mb k,i}(\mb r)\rho J_{\mb k,i}^\dagger (\mb r)e^{-i\mb k\cdot\mb R}-\frac{1}{2}\{J^\dagger_{\mb k,i}(\mb r)J_{\mb k,i}(\mb r),\rho\}\right]
\end{equation}
As for rotation symmetry, described by the unitary $U_{\mbs\theta}=e^{i\mbs\theta\cdot (\mb L_1+\mb L_2)}$, we have
\begin{equation}
\label{eq:rot1}
\begin{aligned}
    U_{\mbs \theta}\mathcal{D}(\rho )U_{\mbs \theta}^\dagger=&\sum_i \int d^3\mb k \left[ \, e^{i\mb k\cdot \Omega_{\mbs \theta}\mb R}J_{\mb k,i }(\Omega_{\mbs \theta}\mb r)U_{\mbs \theta}\rho U_{\mbs \theta}^\dagger J^\dagger_{\mb k,i}(\Omega_{\mbs \theta}\mb r)e^{-i\mb k\cdot \Omega_{\mbs \theta}\mb R}-\frac{1}{2}\{J^\dagger_{\mb k,i}(\Omega_{\mbs \theta}\mb r)J_{\mb k,i}(\Omega_{\mbs \theta}\mb r),\rho \}\right]\\
    =& \sum_i\int d^3\mb k\,  \left[e^{i\mb k\cdot \mb R}J_{\Omega_{\mbs\theta}\mb k,i}(\Omega_{\mbs \theta}\mb r)U_{\mbs \theta}\rho U_{\mbs \theta}^\dagger J^\dagger_{\Omega_{\mbs\theta}\mb k,i}(\Omega_{\mbs \theta}\mb r)e^{-i\mb k\cdot \mb R}-\frac{1}{2}\{J^\dagger_{\Omega_{\mbs\theta}\mb k,i}(\Omega_{\mbs \theta}\mb r)J_{\Omega_{\mbs\theta}\mb k,i}(\Omega_{\mbs \theta}\mb r),\rho\}\right] \, .
    \end{aligned}
\end{equation}
Here, $\Omega_{\mbs\theta}$ represents the rotation matrix associated with angles $\mbs\theta$. Also, in going from the first to the second line, we performed the change of variable $\mb k\rightarrow \Omega_{\mbs\theta}\mb k$. To satisfy $U_{\mbs \theta}\mathcal{D}(\rho )U_{\mbs \theta}^\dagger=\mathcal{D}(U_{\mbs \theta}\rho U_{\mbs \theta}^\dagger)$, we must require

\begin{equation}
\label{eq:rot_inv}
   J_{\Omega_{\mbs\theta}\mb k,i}(\Omega_{\mbs \theta}\mb r)=J_{\mb k,i}(\mb r) \, \quad \forall \, \mbs \theta ,
\end{equation}
implying that $J_{\mb k,i}(\mb r)$ can only be a function of $|\mb k|,|\mb r|$ and $\mb k\cdot \mb r$.

We now impose the ``Ehrenfest theorem'' condition. Using the master equation and comparing to \eqref{eq:classicallimit}, we obtain
\begin{equation}
    \begin{aligned}
         \angbracket{\dt \mb x_a}=&{\mathrm {Tr}}[\dot \rho \mb x_a]={\mathrm {Tr}}[-i[H,\rho]\mb x_a+\mathcal{D}(\rho)\mb x_a]=\angbracket{\frac{\mb p_a}{m_a}} \, ,\\
         \angbracket{\dt \mb p_a}=&{\mathrm {Tr}}[\dot \rho \mb p_a]=\mathrm{Tr}[-i[H,\rho]\mb p_a+\mathcal{D}(\rho)\mb p_a]=\sum_{b\neq a}\angbracket{\frac{d}{d\mb x_a}\frac{G_Nm_am_b}{|\x_a-\x_b|}} \, ,
    \end{aligned}
\end{equation}
Due to the presence of the kinetic energy term in the Hamiltonian, the first line implies $\mathrm{Tr}[\mathcal{D}(\rho)\mb x_a]=0$, which is trivially satisfied, given that the jump operators depend only on the position of the two particles. The implications of the average momentum evolution conditions depend on whether the Newton potential is present in the Hamiltonian or not. In the first case, we have 
\begin{equation}
\label{eq:jump_op_cond}
    \sum_i\int d^3\mb k \,\, \mb k[J_{\mb k,i}(\mb r)J^\dagger_{\mb k,i}(\mb r)]=0 \, ,\qquad \sum_i\int d^3\mb k \left[\frac{\partial J^\dagger_{\mb k,i}(\mb r)}{\partial \mb r}J_{\mb k,i}(\mb r)-J^\dagger_{\mb k,i}(\mb r)\frac{\partial J_{\mb k,i}(\mb r)}{\partial \mb r}\right]=0 \, .
\end{equation}
while in the latter
\begin{equation}
\label{eq:jump_op_cond_2}
    \sum_i\int d^3\mb k \,\, \mb k[J_{\mb k,i}(\mb r)J^\dagger_{\mb k,i}(\mb r)]=0 \, ,\quad \frac{i}{2}\sum_i\int d^3\mb k \left[\frac{\partial J^\dagger_{\mb k,i}(\mb r)}{\partial \mb r}J_{\mb k,i}(\mb r)-J^\dagger_{\mb k,i}(\mb r)\frac{\partial J_{\mb k,i}(\mb r)}{\partial \mb r}\right]=\frac{\partial }{\partial  \mb r} \frac{G_Nm_1m_2}{r} \, .
\end{equation}
Summarizing, the expression for the general form of our quantum channel is thus
\begin{equation}
\label{eq:lind_fin_appA1}
\begin{aligned}
    \mathcal{L}(\rho)=&-i\left[H,\rho \right]
    +\sum_i\int d^3\mb k \left[e^{i\mb k\cdot \mb R}J_{\mb k,i}(\mb r)\rho \,  J^\dagger _{\mb k,i}(\mb r)e^{-i\mb k\cdot \mb R}-\frac{1}{2}\{J^\dagger_{\mb k,i}(\mb r)J_{\mb k,i}(\mb r),\rho\}\right] \, ,
    \end{aligned}
\end{equation}
with $H$ given by either \eqref{eq:ham_initial} or \eqref{eq:ham_initial_2} and the Lindblad operators satisfying either \eqref{eq:jump_op_cond} or \eqref{eq:jump_op_cond_2}, respectively.

\subsection{A subclass of models and approximations in table-top experiments}
\label{app:model_restriction}

As discussed in the main text, we focus primarily on models where the Newton potential is present in the Hamiltonian and we also work with a subclass of dissipation terms defined by the simpler Lindblad evolution shown in Eq.~\eqref{eq:lind_restr}. This is a subset of models contained in our general parametrization defined by Eq.~\eqref{eq:diss_main}. Nevertheless all of the specific alternative gravity models we analyze in Sec. \ref{sec:examples} can be characterized in this simpler form.

This class of models contains both \emph{local} Lindblad operators, which depend on single-particle coordinates, and \emph{non-local} Lindblad operators, depending on the relative position of the particles.

To obtain a parametrization of the local Lindblad operators, we choose

\begin{equation}
\label{eq:J_local_app}
    J_{\mb k,\textrm{loc}}(\mb r)=\sqrt{f_1(k)}e^{i\frac{m_2}{m_1+m_2}\mb k\cdot \mb r}+\sqrt{f_2(k)}e^{-i\frac{m_1}{m_1+m_2}\mb k\cdot \mb r} \, ,
\end{equation}
where $f_{1,2}(k)$ are positive real functions, invariant under rotations. With this parametrization, it can be shown that the Ehrenfest conditions in Eq. \eqref{eq:jump_op_cond} are automatically satisfied. 
We refer to this ansatz as local because the corresponding jump operator can be written as a sum of terms acting separately at the positions $\mathbf{x}_1$ and $\mathbf{x}_2$ of the two particles, namely $L_{\mb k, {\rm loc}}=\sqrt{f_1(k)}e^{i\mb k\cdot\mb x_1}+\sqrt{f_2(k)}e^{i\mb k\cdot\mb x_2}$.
Using the general expression \eqref{eq:lind_fin_appA1}, it is straightforward to show that the local dissipation term is of the form
\begin{equation}
    \mathcal{D}_{loc}(\rho)=\sum_{a,b=1,2}\int d^3\mb k \, \sqrt{f_a(k)f_b(k)} \left(e^{i\mb k\cdot \mb x_a}\rho e^{-i\mb k\cdot \mb x_b}-\frac{1}{2}\{e^{i\mb k\cdot(\mb x_a-\mb x_b)},\rho \}\right)
\end{equation}
To parametrize the non-local Lindblad operators, we set
\begin{equation}
    J_{\mb k,\textrm{non-loc}}(\mb r)= \sqrt{f(k)} J( r) \, ,
\end{equation}
where $f(k)$ is a function of the modulus of $\mb k$, sharply peaked around $k=0$, and $J(r)$ is a function of the modulus of $\mb r$. 
In other words, this minimal non-local ansatz takes $J_{\mathbf{k}}(\mathbf{r})$ with negligible transfer of relative momentum, i.e.\ $\mathbf{k}\simeq 0$. This isolates dissipative processes that act jointly on the two-body system through the relative coordinate while leaving their momenta essentially unaffected.
This allows for the center-of-mass dependence to disappear in \eqref{eq:lind_fin_appA1}, so the non-local dissipation term take the form
\begin{equation}
    \mathcal{D}_{\textrm{non-loc}}(\rho)=J(r)\rho J^\dagger( r)-\frac{1}{2}\{J^\dagger( r)J(r),\rho\} \, ,
\end{equation}
where $(\int{\rm d}^3 \mb k\, f(k))^{1/2}$ is absorbed in $J(r)$.

As also emphasized in the main text, we can make some further simplifications by restricting the motion of the two particles to only 1 spatial dimension, along which the gravitational force is acting. Without loss of generality, we can thus set the $y-$ and $z-$ components of the position operators of the two particles to $0$. Moreover, in the table-top experiments considered in this paper, the displacement of the two particles is much smaller than their initial equilibrium distance $d$. Therefore, we set
\begin{equation}
\label{eq:1dapporx_app}
    x_1=-\frac{d}{2}+\delta x_1,\qquad x_2=\frac{d}{2}+\delta x_2, \qquad \frac{\braket{\delta x_a}}{d}\ll 1 \, .
\end{equation}
For the sake of notational simplicity, here and in the main text we continue to denote the particle displacements by $x_1$ and $x_2$, keeping in mind that these refer to displacements from the equilibrium position.

Let us now analyze how these approximations affect the form of our dissipation terms. The local part of the dissipation can be written as
\begin{equation}
\begin{aligned}
    \mathcal{D}_{\textrm{loc}}(\rho)=&\sum_{a=1,2} \int d^3\mb k \, f_a(k) \left(e^{ik_x x_a}\rho e^{-ik_xx_a} -\rho\right)\\
    +&\int d^3\mb k\,  \sqrt{f_1(k)f_2(k)} e^{ik_x d }\left(e^{ik_xx_1}\rho e^{-ik_xx_2}+e^{ik_xx_2}\rho e^{-ik_xx_1}-e^{ik_x(x_1-x_2)}\rho-\rho e^{ik_x(x_1-x_2)} \right)
    \end{aligned}
\end{equation}
In the second line, given the presence of the $e^{ik_xd}$ term, we can expand the $e^{ik_xx_a}$ terms in series, up to second order. Using the fact that the $f_I(k)$ functions are rotationally invariant, after some algebra we find
\begin{equation}
    \mathcal{D}_{\textrm{loc}}(\rho)=\sum_{a=1,2}\int d^3\mb k \, f_a(k) \left(e^{ik_x x_a}\rho e^{-ik_xx_a} -\rho\right)+\tilde \beta[x_1,[\rho,x_2]]\, ,
\end{equation}
with 
\begin{equation}
   \tilde  \beta=\int d^3\mb k \, k_x^2 \,e^{ik_xL}\sqrt{f_1(k)f_2(k)} \, .
\end{equation}
As for the non-local part, we expand the function $J(r)$ up to second order around the equilibrium distance $d$. According to our approximation \eqref{eq:1dapporx_app}, the relative distance is understood as $r=d+\delta r$, and for notational simplicity we relabel $\delta r$ as $r$. The expansion reads
\begin{equation}
    J(r)\approx j_0+ j_1 r+ j_2r^2 \, .
\end{equation}
where 
Here, the $j_0,j_1,j_2$ coefficients can be complex, in general.
The Ehrenfest conditions \eqref{eq:jump_op_cond} imply that
\begin{equation}
    j_0j_1^*-j_0^*j_1=0\,, \qquad j_2^*j_0-j_0^*j_2=0\,, \qquad j_1j_2^*-j_1^*j_2=0\, .
\end{equation}
As a result, the non-local part of the dissipation can be written as
\begin{equation}
   \mathcal{D}_{\textrm{non-loc}}(\rho) =\frac{|j_1|^2}{2}[r,[\rho,r]]=\frac{|j_1|^2}{2}\left([x_1,[\rho,x_1]]+[x_2,[\rho,x_2]]-2[x_1,[\rho,x_2]]\right)
\end{equation}
The overall dissipation thus takes the form
\begin{equation}
\label{eq:less_gen_diss}
    \mathcal{D}(\rho)=\sum_{a=1,2}\int d^3\mb k f_a(k)\left(e^{ik_xx_a}\rho e^{-ik_xx_a}-\rho\right)+\alpha[x_1,[\rho,x_1]]+\alpha[x_2,[\rho,x_2]]+\beta [x_1,[\rho,x_2]] \, ,
\end{equation}
where we defined $\alpha=|j_1|^2/2$ and $\beta=\tilde\beta-|j_1|^2$. 
Notice that the double-commutator terms involving single particle coordinates can be written in the form
\begin{equation}
    \alpha[x_a,[\rho,x_a]]=\int d^3\mb k\, g(k)\left(e^{ik_xx_a}\rho e^{-ik_xx_a}-\rho\right), \qquad g(k)=-\alpha \partial_{k_x}^2\delta^3(\mb k) \, ,
\end{equation}
so they can be formally reabsorbed in a redefinition of $f_I(k)$. In this approximation, we also expand the Newton potential about the equilibrium distance $d$, up to second order in the small displacements
\begin{equation}
    -\frac{G_Nm_1m_2}{|\mb x_1-\mb x_2|}\sim -\frac{G_Nm_1m_2}{d}+\frac{G_Nm_1m_2}{d^2}(x_1-x_2)-\frac{G_Nm_1m_2}{d^3}(x_1-x_2)^2 \, .
\end{equation}
Absorbing the linear term in a redefinition of the equilibrium position and neglecting constant terms, we obtain the final form of our quantum channel 
\begin{equation}
\label{eq:lind_restr_app}
\begin{aligned}
    \mathcal{L}(\rho)=&-i\left[\sum_{a=1}^2 \frac{\mb p_a^2}{2m_a}-\frac{G_Nm_1m_2}{d^3}(x_1-x_2)^2,\rho \right]
    \\
    &+\sum_{a=1,2}\int d^3\mb k\, f_a(k)\left(e^{ik_xx_a}\rho e^{-ik_xx_a}-\rho\right)+\beta [x_1,[\rho,x_2]] \, .
    \end{aligned}
\end{equation}
Of course, based on the physical scenario under study, the Hamiltonian may contain further non-gravitational terms, as explained below Eq. \eqref{eq:lind_restr}.
The channel \eqref{eq:lind_restr_app} depends on two positive and isotropic real functions $f_a(k)$ and a real parameter $\beta$. We identify the first two integral terms as sources of local noise and the double-commutator term as non-local noise, given that it depends on the coordinates of both particles. 

For completeness, we also report the analogous of \eqref{eq:lind_restr_app} without restricting to one spatial dimension. Expanding the relative position around the equilibrium distance $d$, for the local part of the dissipation we have
\begin{equation}
\begin{aligned}
    \mathcal{D}_{\textrm{loc}}(\rho)=&\sum_{a=1,2} \int d^3\mb k \, f_a(k) \left(e^{i\mb k\cdot \mb x_a }\rho e^{-i\mb k\cdot \mb x_a} -\rho\right)+\tilde{\beta}_{ij}[x_1^i,[\rho,x_2^j]]
    \end{aligned}
\end{equation}
with
\begin{equation}
    \tilde{\beta}_{ij}=\int d^3 \mb k  \, \,k_ik_j\sqrt{f_1(k)f_2(k)}e^{i\mb k\cdot \mb d} 
\end{equation}
The expansion for the non-local Lindblad operator $J(r)$ generalizes to 
\begin{equation}
    J(r)\approx j_0+j_{1,i} r^i + j_{2,ij} r^ir^j
\end{equation}
with $j_0,j_{1,i},j_{2,ij}$ complex, in general. The Ehrenfest conditions \eqref{eq:jump_op_cond} imply that
\begin{equation}
    j_{1,i}^*j_0=j_0^*j_{1,i} \,,\, \qquad j_{i,i}^*j_{1,j}+2j_0j_{2,ij}^*=j_{1,i}j_{1,j}^*+2j_0^*j_{2,ij}\,,\qquad j^*_{1,i} j_{2,jk}+2j_{2,ij}^*j_{1,k}=2j_{2,ij}j_{1,k}^*+j_{1,i}j_{2,jk}^* \, .
\end{equation}
The non-local dissipation term takes the form
\begin{equation}
\begin{aligned}
    \mathcal{D}_{\mathrm{non-loc}}(\rho)=&j_{1,i}j_{1,j}^*\left(r^i\rho r^j-\frac{1}{2}r^ir^j\rho -\frac{1}{2}\rho r^ir^j\right)\\
    =&j_{1,i}j_{1,j}^*\sum_a\left(x_a^i\rho x_a^j-\frac{1}{2}x_a^ix_a^j\rho -\frac{1}{2}\rho x_a^ix_a^j\right)- j_{1,i}j_{1,j}^*\left(x_{1}^i\rho x_2^j + x_2^i\rho x_1^j -x_1^ix_2^j\rho - \rho x_1^ix_2^j \right) \, .
    \end{aligned}
\end{equation}
It can also be checked that the single particle terms coming from $\mathcal{D}_{\mathrm{non-loc}}(\rho)$ can be rewritten as
\begin{equation}
    j_{1,i}j_{1,j}^*\left(x_a^i\rho x_a^j-\frac{1}{2}x_a^ix_a^j\rho -\frac{1}{2}\rho x_a^ix_a^j\right)=\int d^3 \mb k \, g(k)\left(e^{i\mb k \cdot \mb x_a}\rho e^{-i\mb k \cdot \mb x_a}-\rho\right) \, , \qquad g(k)  = -\frac{j_{1,i}j_{1,j}^*}{2}\partial_{k_i}\partial_{k_j}\delta^3(\mb k) \, .
\end{equation}
In three dimensions, the expansion of the Newtonian potential around the equilibrium distance $d$ is
\begin{equation}
-\frac{G_N m_1 m_2}{|\mathbf{x}_1 - \mathbf{x}_2|}
\;\sim\;
-\frac{G_N m_1 m_2}{d}
+ \frac{G_N m_1 m_2}{d^{3}}\,d_i\,(x_1 - x_2)^{i}
- \frac{G_N m_1 m_2}{2d^5}\,(d_id_j-\delta_{ij}d^2)\,(x_1 - x_2)^{i} (x_1 - x_2)^{j}
\end{equation}
Neglecting the constant term, we arrive at the final form of our channel in three dimensions
\begin{equation}
\label{eq:lind_restr_app_3D}
\begin{aligned}
    \mathcal{L}(\rho)=&-i\left[\sum_{a=1}^2 \frac{\mb p_a^2}{2m_a}+ \frac{G_N m_1 m_2}{d^{3}}\,d_i\,(x_1 - x_2)^{i}-\frac{G_Nm_1m_2}{2d^5}(d_id_j-\delta_{ij}d^2)(x_1-x_2)^i(x_1-x_2)^j,\rho \right]
    \\
    &+\sum_{a=1,2}\int d^3\mb k\, f_a(k)\left(e^{i\mb k\cdot \mb x_a}\rho e^{-i\mb k\cdot \mb x_a}-\rho\right)+\beta_{ij} \left(x_1^i\rho x_2^j + x_2^i\rho x_1^j - x_1^ix_2^j\rho -\rho x_1^ix_2^j\right) \, 
    \end{aligned}
\end{equation}
where we defined $\beta_{ij}=\tilde\beta_{ij}-j_{1,i}j_{1,j}^*$.

\section{Entanglement criteria}
\label{app:ent_criteria}

\subsection{Two mechanical oscillators}
\label{app:two_oscillators}
In this appendix, we prove the noise inequality in Eq.~\eqref{eq:ent_cond_oscillators}. However, unlike the main text where we used the simplified Lindbladian of Eq.~\eqref{eq:diss_main}, here we use the fully general form of Eq.~\eqref{eq:lind_restr}, to emphasize that these bounds work without any of the extra assumptions made in Appendix~\ref{app:model_restriction}. We will also show that the noise inequality in  Eq.~\eqref{eq:ent_cond_oscillators} holds for models where the Newtonian interaction is sourced by the dissipation terms, as anticipated in the main text.

Starting from the definition in Eq. \eqref{eq:covmatdef}, a straightforward computation shows that the time evolution for the covariance matrix is
    \begin{equation}
\label{eq:covmateom}
    \dot \gamma = x^T\gamma +\gamma x + y \, ,
\end{equation}
where $x$ is a $4\times 4$ matrix depending on the coefficients of the quadratic terms of the Hamiltonian:  
\begin{equation}
    x=-X\Delta_2 \, , \qquad  H=\frac{1}{2}X_{ij}M_iM_j\, , 
\end{equation}
and $y$ is also a $4\times 4$ matrix depending on the non-unitary part of the dynamics through
\begin{equation}
    y_{ij}=\mathrm{Tr}\left[\mathcal{D}(\rho)(M_iM_j+M_jM_i)\right]-2\mathrm{Tr}\left[\mathcal{D}(\rho)M_i\right]\mathrm{Tr}\left[\rho M_j\right]-2\mathrm{Tr}\left[\rho M_i\right]\mathrm{Tr}\left[\mathcal{D}(\rho)M_j\right] \, .
\end{equation}
Since we will be interested in the evolution of the system for short times, we can write
\begin{equation}
\label{eq:cov_mat_first_ord}
    \gamma(t)\approx \gamma_0 +{t}(x^T\gamma_0+\gamma_0x+y) \,,
\end{equation}
with $\gamma_0=\gamma(0)$, and the components of $x$ and $y$ will be evaluated at $t=0$. We assume that our two masses are prepared in their ground states, so the initial covariance matrix will be of the form
\begin{equation}
\label{eq:initial_state_app}    \gamma_0=\textrm{diag}\left[1,1,1,1\right] \, .
\end{equation}
We begin by analyzing the case in which the Newton potential is in the Hamiltonian. Using the expression for the Hamiltonian in  \eqref{eq:lind_restr_app} with the addition of harmonic potential terms $\sum_am_a\omega_a^{'2}/2$ and the general dissipation \eqref{eq:diss_main}, we have
\begin{equation}
\label{eq:X_and_g}
    X=\begin{pmatrix}
        \omega_1 & 0 & 2g & 0\\
        0 & \omega_1 & 0 & 0 \\
        2g & 0 & \omega_2 & 0 \\
        0 & 0 & 0 & \omega_2
    \end{pmatrix}\,, \qquad g=\frac{\alpha_G}{p_{1,0}p_{2,0}} \, ,\quad  \omega_a^2=\omega_a^{'2}-\frac{2\alpha_G}{m_a}\, ,
\end{equation}
with $p_{0,a}=\sqrt{m_a\omega_a}$. As for the non-unitary contribution, the matrix $y$ is 
\begin{equation}
    y=2\begin{pmatrix}
        0 & 0 & 0 & 0\\
        0 & \Gamma_1 & 0 & \Gamma_{12}\\
        0 & 0 & 0 & 0\\
        0 & \Gamma_{12} & 0 & \Gamma_2
    \end{pmatrix} \, ,
\end{equation}
where $\Gamma_a,\Gamma_{12}$ are the time derivatives for the excess variance of the momenta, which can be calculated using \eqref{eq:diss_main}:
\begin{equation}
\label{eq:delta_defs}
\begin{aligned}
    p_{1,0}^2\Gamma_1
    ={\rm tr}[\mathcal{D}(\rho) p_1^2]_{t=0}=\frac{d\angbracket{p_1^2}_e}{dt}\Bigg|_{t=0}
    &=
    \sum_i\int d^3\mb k\,
    \left\langle
    \mathcal{J}_{1,\mb k,i}^\dagger \mathcal{J}_{1,\mb k,i}
    \right\rangle ,
    \\
    p_{2,0}^2\Gamma_2
    ={\rm tr}[\mathcal{D}(\rho) p_2^2]_{t=0}=\frac{d\angbracket{p_2^2}_e}{dt}\Bigg|_{t=0}
    &=
    \sum_i\int d^3\mb k\,
    \left\langle
    \mathcal{J}_{2,\mb k,i}^\dagger \mathcal{J}_{2,\mb k,i}
    \right\rangle ,
    \\
    p_{1,0}p_{2,0}\Gamma_{12}
    ={\rm tr}[\mathcal{D}(\rho) p_1p_2]_{t=0}
    =\frac{d\angbracket{p_1p_2}_e}{dt}\Bigg|_{t=0}&=
    -\sum_i\int d^3\mb k\,
    {\rm Re}\left[
    \left\langle
    \mathcal{J}_{1,\mb k,i}^\dagger \mathcal{J}_{2,\mb k,i}
    \right\rangle
    \right] ,
\end{aligned}
\end{equation}
with
\begin{equation}
\label{eq:Jcal_defs}
\mathcal{J}_{1,\mb k,i}
:=
\frac{\partial J_{\mb k,i}}{\partial r_x}
+\frac{i m_1 k_x}{M}J_{\mb k,i},
\qquad
\mathcal{J}_{2,\mb k,i}
:=
\frac{\partial J_{\mb k,i}}{\partial r_x}
-\frac{i m_2 k_x}{M}J_{\mb k,i}.
\end{equation}
Notice that the size of coefficient $\Gamma_{12}$ is bounded by the sizes of coefficients $\Gamma_1\Gamma_2$, reflecting the positivity of our quantum channel. Indeed, for each fixed $(\mb k,i)$, 
\begin{equation}
\left|
{\rm Re}\left[
\left\langle
\mathcal{J}_{1,\mb k,i}^\dagger \mathcal{J}_{2,\mb k,i}
\right\rangle
\right]
\right|
\le
\left|
\left\langle
\mathcal{J}_{1,\mb k,i}^\dagger \mathcal{J}_{2,\mb k,i}
\right\rangle
\right|
\le
\sqrt{
\left\langle
\mathcal{J}_{1,\mb k,i}^\dagger \mathcal{J}_{1,\mb k,i}
\right\rangle
\left\langle
\mathcal{J}_{2,\mb k,i}^\dagger \mathcal{J}_{2,\mb k,i}
\right\rangle
}.
\end{equation}
Taking the sum and integral, the left hand side becomes $|p_{1,0}p_{2,0}\Gamma_{12}|$.
One more application of the Cauchy--Schwarz inequality is needed, this time for the sum/integral itself, namely
$\sum_\alpha \sqrt{X_\alpha Y_\alpha}\le \sqrt{(\sum_\alpha X_\alpha)(\sum_\alpha Y_\alpha)}$
with $\alpha=(\mb k,i)$. This gives
\begin{equation}
|p_{1,0}p_{2,0}\Gamma_{12}|^2
\le
\left(
\sum_i\int d^3\mb k\,
\left\langle
\mathcal{J}_{1,\mb k,i}^\dagger \mathcal{J}_{1,\mb k,i}
\right\rangle
\right)
\left(
\sum_i\int d^3\mb k\,
\left\langle
\mathcal{J}_{2,\mb k,i}^\dagger \mathcal{J}_{2,\mb k,i}
\right\rangle
\right),
\end{equation}
which is precisely
\begin{equation}
\Gamma_{12}^2 \le \Gamma_1 \Gamma_2.
\label{eq:gamma12 condition}
\end{equation}
Notably, these rates reduce to much simpler expressions when the dissipation is of the form \eqref{eq:lind_restr} :
\begin{equation}
    p_{a,0}^2\Gamma_a=\int d^3\mb k \, k_x^2\, f_a(k) \ , \qquad p_{1,0}p_{2,0}\Gamma_{12}=\beta \, .
\end{equation}
Using the general expressions for $x$ and $y$, and with the initial state \eqref{eq:initial_state_app}, three of the four eigenvalues of $\tilde\gamma+i\Delta_2=K\gamma K+i\Delta_2$, with $K=\textrm{diag}[1,1,1-1]$, can be shown to be positive definite, while the fourth is 
\begin{equation}
\label{eq:eigenvalue_oscillators}
    \lambda=\frac{t}{2}\left(\Gamma_1+\Gamma_2-\sqrt{(\Gamma_1-\Gamma_2)^2+4\Gamma_{12}^2+16g^2}\right)
\end{equation}
which is negative if
\begin{equation}
\Gamma_1\Gamma_2<\Gamma_{12}^2+4g^2
\end{equation} 
Recalling definitions  \eqref{eq:delta_defs} and \eqref{eq:X_and_g}, this is equivalent to 
\begin{equation}
\label{eq:var_cond_app}
    \frac{d\angbracket{p_1^2}_e}{dt}\Bigg|_{t=0}\frac{d\angbracket{p_2^2}_e}{dt}\Bigg|_{t=0}<\left(\frac{d\angbracket{p_1p_2}_e}{dt}\Bigg|_{t=0}\right)^2+ 4\alpha_G^2
\end{equation}

Here, the rates of the variance of the momenta are evaluated at $t=0$ in virtue of our approximation in \eqref{eq:cov_mat_first_ord}. We now claim that \eqref{eq:var_cond_app} is also satisfied if 
\begin{equation}
    \label{eq:noise_bound_app}\frac{d\angbracket{p_1^2}_e}{dt}\Bigg|_{t=0}+\frac{d\angbracket{p_2^2}_e}{dt}\Bigg|_{t=0}<4\alpha_G \, .
\end{equation}
Indeed, using the arithmetic mean-geometric mean inequality $(a+b)/2>\sqrt{|ab|}$, for positive $a,b$:
\begin{equation}
    \frac{d\angbracket{p_1^2}_e}{dt}\Bigg|_{t=0}\frac{d\angbracket{p_2^2}_e}{dt}\Bigg|_{t=0}<\frac{1}{4}\left(\frac{d\angbracket{p_1^2}_e}{dt}\Bigg|_{t=0}+\frac{d\angbracket{p_2^2}_e}{dt}\Bigg|_{t=0}\right)^2<4\alpha_G^2 \, ,
\end{equation}
so \eqref{eq:var_cond_app} is certainly satisfied.

In the case where the Newton potential is absent from the Hamiltonian, the expression for $X$ is given by
\begin{equation}
    X=\begin{pmatrix}
        \omega_1 & 0 & 0 & 0\\
        0 & \omega_1 & 0 & 0 \\
        0 & 0 & \omega_2 & 0 \\
        0 & 0 & 0 & \omega_2 
    \end{pmatrix} \,, \quad  \omega_a^2=\omega_a^{'2}\,
\end{equation}
while $y$ is
\begin{equation}
    y=2\begin{pmatrix}
        0 & -g_1 & 0 &   g\\
        -g_1 & \Gamma_1 & g & \Gamma_{12}\\
        0 & g & 0 & -g_2\\
        g & \Gamma_{12} & -g_2 & \Gamma_2
    \end{pmatrix} \, , \quad g_a=\frac{\alpha_G}{p_{a,0}^2}\,  ,\quad g=\frac{\alpha_G}{p_{1,0}p_{2,0}} \, ,
\end{equation}
with the $\Gamma_a,\Gamma_{12}$ terms being defined as in \eqref{eq:delta_defs}. In this case, contributions coming from $x_ap_b$ terms are non-zero and proportional to $\alpha_G$, due to condition \eqref{eq:jump_op_cond_2}. For example, consider
\begin{equation}
\begin{aligned}
    \dot \gamma_{12}(0)&= \sum_{i}\int d^3\mb k \, \angbracket{J_{\mb k,i}^\dagger e^{-ik_x R_x}x_1p_1 e^{i k_xR_x}J_{\mb k,i}-\frac{1}{2}\{J^\dagger_{\mb k,i} J_{\mb k,i},x_1p_1\}}_0\\
    =&\frac{i}{2}\sum_i\int d^3\mb k \angbracket{x_1\left(\frac{\partial J_{\mb k,i}^\dagger}{\partial x_1}J_{\mb k,i}-J_{\mb k,i}\frac{\partial J_{\mb k,i}}{\partial x_1}\right)}_0\\
    =&G_Nm_1m_2\angbracket{x_1\frac{\partial }{\partial x_1}\left(-\frac{1}{d}+\frac{(x_1-x_2)}{d^2}-\frac{(x_1-x_2)^2}{d^3}\right)}_0\\
    =&-\frac{2G_Nm_1m_2}{d^3}\angbracket{x_1(x_1-x_2)}_0=-\frac{G_Nm_1m_2}{d^3}x_{1,0}^2=-\frac{\alpha_G}{p_{1,0}^2} \, .
    \end{aligned}
\end{equation}
As a consistency check, from the second line notice that this term would have been $0$ if the Hamiltonian were already present in the Hamiltonian in virtue of 
\begin{equation}
    \sum_i\int d^3\mb k\left(\frac{\partial J_{\mb k,i}^\dagger}{\partial x_1}J_{\mb k,i}-J_{\mb k,i}\frac{\partial J_{\mb k,i}}{\partial x_1}\right)=0 \, .
\end{equation}
Calculations involving other $x_ap_b$ contributions are similar in style. Computing the eigenvalues of $\tilde\gamma+i\Delta_2=K\gamma K+i\Delta_2$, we find once again that three of them are positive definite, while the fourth is identical to \eqref{eq:eigenvalue_oscillators}, so the entangling condition is the same as we found in the previous case.

\subsection{Masses with two position states}
\label{app:two_qubits}
For the experimental setup involving two particles prepared in a superposition $\delta x$, described in Sec.~\ref{sec:qubits}, the position operators can be described by $z$-Pauli matrices and the Hamiltonian can be written as
\begin{equation}
\label{eq:two_qubit_ham_app}
    H\approx \frac{\alpha_G\delta x^2}{2}\sigma_1^z\sigma_2^z \, .
\end{equation}
Here, we have neglected constant contributions to the Hamiltonian and the kinetic energy terms. This last approximation is valid for sufficiently heavy masses, such as those in~\cite{Bose:2017nin}.

Restricting our attention to the class of models described by Eq. \eqref{eq:lind_restr_app}, and using $e^{i\theta\sigma^z}=\cos(\theta)+i\sin(\theta)\sigma^z$, the dissipative part of the dynamics can be written as
\begin{equation}
\label{eq:diss_qubits_app}
    \begin{aligned}
    \mathcal{D}(\rho)=&\sum_{a=1,2}\Gamma_a(\sigma_a^z\rho\sigma_a^z-\rho)
    +\frac{\beta\delta x^2}{4}(\sigma_1^z\rho\sigma_2^z+\sigma_2^z\rho\sigma_1^z-\sigma_1^z\sigma_2^z\rho-\rho\sigma_1^z\sigma_2^z) \, ,\\
    \end{aligned}
\end{equation}
with 
\begin{equation}
\label{eq:gamma_qubits_app}
    \Gamma_a=\int d^3\mb k \, f_a(k)\sin^2\left(\frac{k_x\delta x}{2}\right) \, .
\end{equation}

Denoting the components of the density matrix by $\rho_{ijhk}=\bra{ij}\rho\ket{hk}$, with $i,j,k,l$ corresponding to $+1$ for $\ket{R}$ and to $-1$ for $\ket{L}$, the equation of motion for the density matrix elements is 
\begin{equation}
    \dot \rho_{ijhk}=f_{ijhk}\rho_{ijhk \ ,}
\end{equation}
with 
\begin{equation}
\label{eq:f}
\begin{aligned}
    f_{ijhk}=&-i\frac{\alpha_G\delta x^2}{2}(ij-hk)+\Gamma_1(ih-1)+\Gamma_2(jk-1)+\frac{\beta\delta x^2}{4}(ik+jh-ij-hk) \, ,
    \end{aligned}
\end{equation}
which is solved by
\begin{equation}
\label{eq:rho_sol_qubit}
    \rho_{ijhk}(t)=e^{if_{ijhk}t}\rho_{ijhk}(0) \, .
\end{equation}
For two particles initially prepared in a superposition of left and right states, the initial state of the system is described by 
\begin{equation}
    \rho(0)=\frac{1}{4}(\mathbf{1}+\sigma^x_1)(\mathbf{1}+\sigma^x_2) \, .
\end{equation}
Using solution \eqref{eq:rho_sol_qubit} and computing the eigenvalues of its partial transpose, the negativity at first order in $t$ is
\begin{equation}
    \mathcal{N}(t)\approx \frac{t}{4}\left[\left|\Gamma_1+\Gamma_2-\sqrt{(\Gamma_1-\Gamma_2)^2+\frac{\beta^2\delta x^4}{4}+\alpha_G^2\delta x^4} \right|-(\Gamma_1+\Gamma_2)+\sqrt{(\Gamma_1-\Gamma_2)^2+\frac{\beta^2\delta x^4}{4}+\alpha_G^2\delta x^4}\right] \, ,
\end{equation}
This quantity is greater than $0$ if and only if
\begin{equation}
    \label{eq:qubi_cond_app}\Gamma_1+\Gamma_2<\sqrt{(\Gamma_1-\Gamma_2)^2+\frac{\beta^2\delta x^4}{4}+\alpha_G^2\delta x^4} \, .
\end{equation}
Therefore, entanglement is generated if and only if condition \eqref{eq:qubi_cond_app} is satisfied. 
We now show how \eqref{eq:qubi_cond_app} is equivalent to condition \eqref{eq:bound_qubits_simple} of the main text. For each particle, we defining the visibility as $|\angbracket{\sigma_a^-(t)}|$, where $\sigma_a^-=(\sigma_a^x-\sigma_a^y)/2=\ket{L}\bra{R}_a$, and compute its excess noise in the initial phase of the experiment, using \eqref{eq:diss_qubits_app}:
\begin{equation}
    \left|\frac{d|\angbracket{\sigma_a^-}_e|}{dt}\right|_{t=0}=\Gamma_a \, .
\end{equation}
It simply follows that
\begin{equation}
    \left|\frac{d|\angbracket{\sigma_1^-}_e|}{dt}+\frac{d|\angbracket{\sigma_2^-}_e|}{dt}\right|_{t=0}= \Gamma_1+\Gamma_2<\sqrt{(\Gamma_1-\Gamma_2)^2+\frac{\beta^2\delta x^4}{4}+\alpha_G^2\delta x^4}
\end{equation}
Therefore, if
\begin{equation}
\label{eq:vis_cond_app}
   \left|\frac{d|\angbracket{\sigma_1^-}_e|}{dt}\right|+\left|\frac{d|\angbracket{\sigma_2^-}_e|}{dt}\right|<\alpha_G\delta x^2,
\end{equation}
then relation \eqref{eq:qubi_cond_app} is satisfied and the channel is entangling, which is the result stated in Sec. \ref{sec:qubits}.

As for the positivity condition of the channel in this case, it is useful to note that Eq.~\eqref{eq:diss_qubits_app} is already in the quadratic form,
\begin{equation}
\mathcal D(\rho)
=
\sum_{I,J=1,2} C_{IJ}\left(\sigma_I^z\rho \sigma_J^z-\frac12\{\sigma_J^z \sigma_I^z,\rho\}\right),
\qquad
C=
\begin{pmatrix}
\Gamma_1 & \beta\delta x^2/4\\
\beta\delta x^2/4 & \Gamma_2
\end{pmatrix}.
\end{equation}
Complete positivity of the corresponding Markovian evolution requires \(C\ge0\). Since \(C\) is a real symmetric \(2\times2\) matrix and $\Gamma_1, \Gamma_2\ge0$, this implies 
$\det C\ge0$, leading to
\begin{equation}
\label{eq:beta_bound_qubit_app}
16\,\Gamma_1\Gamma_2\ge\beta^2\delta x^4.
\end{equation}
This condition can be rewritten as
\begin{equation}
(\Gamma_1+\Gamma_2)^2\ge(\Gamma_1-\Gamma_2)^2+\frac14 \beta^2\delta x^4,
\end{equation}
so that Eq.~\eqref{eq:qubi_cond_app} cannot be satisfied
without the gravitational coupling $\alpha_G=0$. Therefore the \(\beta\)-term alone does not generate initial negativity.

\subsection{Mechanical mass coupled to two-state system}
\label{app:hybrid}
For the hybrid system of Sec. \ref{sec:osc_qubit}, we find it convenient to describe the oscillator (particle 1) degrees of freedom in terms of creation and annihilation operators satisfying $[a,a^\dagger]=1$. We will model an atom (particle 2) of the atom interferometer as a two-position state system, as we have done in Appendix \ref{app:two_qubits}, so the position operator can be written as $x_2=\delta x\sigma^z/2$, where $\delta x$ is the superposition width. In terms of $a,a^\dagger$, the position and momentum of the oscillator are
\begin{equation}
    x_1=\frac{(a+a^\dagger)}{\sqrt{2M\omega}} \, , \qquad p_1=i\sqrt{\frac{M\omega}{2}}(a^\dagger -a)\, ,
\end{equation}
and the Hamiltonian of the system is
\begin{equation}
    H=\omega a^\dagger a+g(a+a^\dagger)\sigma_z \, ,\qquad g=\frac{G_NMm\delta x}{L^3\sqrt{2M\omega}}
\end{equation}
while the dissipation \eqref{eq:less_gen_diss} assumes the form
\begin{equation}
\label{eq:hybrid_diss_app}
\begin{aligned}
    \mathcal{D}(\rho)=&\int d^3\mb k \, f_1(k)\left[D\left(\frac{ik_x}{\sqrt{2M\omega}}\right)\rho D^\dagger\left(\frac{ik_x}{\sqrt{2M\omega}}\right)-\rho\right] \\
    +&\Gamma_2 [\sigma^z \rho \sigma^z-\rho]+\frac{\beta\delta x}{2\sqrt{2M\omega}}[(a+a^\dagger),[\rho,\sigma^z]] \, ,
    \end{aligned}
\end{equation}
where the definition of $\Gamma_2$ is the same as in Eq. \eqref{eq:gamma_qubits_app}. 
Here, we have introduced the displacement operators $D(\alpha)=\exp(\alpha a^\dagger-\alpha^* a)$, for complex $\alpha$, whose action on the vacuum is $D(\alpha)\ket{0}=\ket{\alpha}$, where $\ket{\alpha}$ is a coherent state, meaning that $a\ket{\alpha}=\alpha \ket{\alpha}$.

For the initial state of the system, we consider the oscillator in the ground state and the atom in superposition, namely
\begin{equation}
\label{eq:hybrid_initial_state}
    \ket{\psi(0)}=\frac{\ket{L}+\ket{R}}{\sqrt{2}}\ket{0} \rightarrow \rho(0)=\frac{1}{2}(\mathbf{1}+\sigma^x)\ket{0}\bra{0}=\frac{1}{2}\begin{pmatrix}
       \ket{0}\bra{0} && \ket{0}\bra{0}\\
       \ket{0}\bra{0} && \ket{0}\bra{0}
    \end{pmatrix}  \, .
    \end{equation}  

The objective is to determine a sufficient condition for entanglement generation by computing the eigenvalues of the partial transpose of the density matrix. We will do so by working perturbatively, up to first order in $t$. 
At order $0$, taking the partial transpose with respect to the atom, we have $\rho^{T_B}(0)=\rho(0)$, and the eigenvalue equation $\rho^{T_B}(0)\ket{\psi}=\lambda\ket{\psi}$, with
\begin{equation}
    \ket{\psi}=\begin{pmatrix}
        \ket{u} \\\ket{v}
    \end{pmatrix} \, ,
\end{equation}
is solved by $\lambda=1$, corresponding to the eigenvector specified by $\ket{u}=\ket{v}=\ket{0}$. Another degenerate solution is $\lambda=0$, corresponding to the eigenspace spanned by
\begin{equation}
\label{eq:deg_eigenspace}
    \ket{\phi_0}=\frac{1}{\sqrt{2}}\begin{pmatrix}
        -\ket{0}\\\ket
        0
    \end{pmatrix}, \qquad \ket{\phi_+^n}=\frac{1}{\sqrt{2}}\begin{pmatrix}
        \ket{n} \\
        \ket{n}
    \end{pmatrix},\qquad \ket{\phi_-^n}=\frac{1}{\sqrt{2}}\begin{pmatrix}
        \ket{n}\\
        -\ket{n}
    \end{pmatrix}, \qquad n\geq 1
\end{equation}
We now compute the first order corrections to the $\lambda=0$ degenerate eigenvalue, denoted by $\lambda^{(1)}$, and determine under which condition they are negative, yielding our sufficient condition for entanglement.

We determine the evolution of the density matrix up to first order in $t$ : $\rho(t)\approx \rho(0)+Vt$. Then, we construct the numerical matrix $C_{ij}=\bra{v_i}V^{T_B}\ket{v_j}$, obtained by projecting $V^{T_B}$ onto the subspace spanned by the vectors $\ket{v_i}$ in \eqref{eq:deg_eigenspace}. The eigenvalues of the partial transpose of $C_{ij}$ are the desired corrections $\lambda^{(1)}$.

The contribution to $V$ coming from the Hamiltonian is
\begin{equation}
    V\supset -i[H,\rho(0)]t = \frac{igt}{2}\begin{pmatrix}
        \ket{0}\bra{1}-\ket{1}\bra{0} & -\ket{0}\bra{1}-\ket{1}\bra{0}\\
        \ket{0}\bra{1}+\ket{1}\bra{0} & \ket{1}\bra{0}-\ket{0}\bra{1}
    \end{pmatrix} \, .
\end{equation}
Taking the partial transpose of this contribution, its only nonzero projection on the subspace generated by \eqref{eq:deg_eigenspace} is
\begin{equation}
\label{eq:hybrid_pert_ham}
    \bra{\phi_+^1}-i[H,\rho(0)]^{T_B}t\ket{\phi_0}=igt\,.
\end{equation}
In the case of no dissipation, it is easy to show that this would result in corrections to the $\lambda=0$ eigenvalue given by $\lambda^{(1)}=\pm gt$, implying entanglement generation.

The contribution to $V$ coming from the oscillator dissipation term is 
\begin{equation}
\begin{aligned}
    V\supset \mathcal{D}_{osc}(\rho(0))t = \frac{t}{2}(\mathbf{1}+\sigma_x)\int d^3\mb k\, f_1( k) \left(\Ket{\frac{ik_x}{\sqrt{2M\omega}}}\Bra{\frac{ik_x}{\sqrt{2M\omega}}}-\ket{0}\bra{0}\right) \, .
    \end{aligned}
\end{equation}
The partial transpose leaves this term invariant. Using the fact that $(\mathbf{1}+\sigma_x)\ket{\phi_0}=(\mathbf{1}+\sigma_x)\ket{\phi_-^n}=0$, the only non-trivial projections onto the subspace of interest are given by
\begin{equation}
\label{eq:hybrid_pert_osc}
    D_{nm}t=\bra{\phi_+^n}\mathcal{D}_{osc}(\rho(0))t\ket{\phi_+^m}=i^{n-m}t\int d^3 \mb k\, f_1(k) \frac{e^{-\frac{k_x^2}{2M\omega}}}{\sqrt{m!n!}} \frac{k_x^{n+m}}{(2M\omega)^{\frac{n+m}{2}}} \, .
\end{equation}

The contribution to $V$ coming from the two-state dissipation term is
\begin{equation}
    V\supset\mathcal{D}_{qub}(\rho(0))t=-\Gamma_2\sigma^x\ket{0}\bra{0}t \, .
\end{equation}
Again, the partial transpose leaves this term invariant. The action of this term on $\ket{\phi_+^n}$ and $\ket{\phi_-^n}$ states gives $0$, since $n\geq 1$. The only nontrivial projection onto the subspace of interest is 
\begin{equation}
\label{eq:hybrid_pert_qubit}
    \bra{\phi_0}\mathcal{D}_{qub}(\rho(0))t\ket{\phi_0}=\Gamma_2t
\end{equation}

As for the mixed dissipation term, we have
\begin{equation}
    V\supset \mathcal{D}_{mix}(\rho(0))t=\frac{-it\beta\delta x}{2\sqrt{2M\omega}}\sigma^y(\ket{0}\bra{1}-\ket{0}\bra{1})
\end{equation}
Applying the partial transpose, the only nontrivial projection onto the subspace of interest is
\begin{equation}
\label{eq:hybrid_pert_mix}
    \bra{\phi_+^1}\mathcal{D}_{mix}(\rho(0))^{T_B}t\ket{\phi_0}=\frac{\beta\delta xt}{2\sqrt{2M\omega}}
\end{equation}
Putting together results \eqref{eq:hybrid_pert_ham},\eqref{eq:hybrid_pert_osc},\eqref{eq:hybrid_pert_qubit},\eqref{eq:hybrid_pert_mix},  the first order corrections to the eigenvalue $\lambda=0$ are the eigenvalues of the matrix
\begin{equation}
    C_{ij}=\bra{v_i}V\ket{v_j}=\begin{pmatrix}
        \Gamma_2 & -ig+\frac{\beta\delta x}{2\sqrt{2M\omega}} & 0  & \cdots \\
        ig+\frac{\beta\delta x}{2\sqrt{2M\omega}} & D_{11} & D_{12} & \cdots \\
        0 & D_{21} & D_{22} & \cdots\\
        \vdots & \vdots & \vdots & \ddots
    \end{pmatrix}t
\end{equation}
written in the basis $\ket{v_i}=\{\ket{\phi_0},\ket{\phi_+^n},\ket{\phi_-^n}\}$. Notice that according to our intermediate results, all matrix elements involving $\ket{\phi_-^n}$ are zero. 

The eigenvalue equation $C\mb a=\lambda^{(1)} t\mb a$, with $\mb a=(a_0,a_1,\dots)$ is equivalent to the following system of equations
\begin{equation}
\label{eq:hybrid_system_1}
\begin{aligned}
    &\Gamma_2 a_0 -\left(ig-\frac{\beta\delta x}{2\sqrt{2M\omega}}\right)a_1=\lambda^{(1)} a_0\\
    &\left(ig+\frac{\beta\delta x}{2\sqrt{2M\omega}}\right)a_0 + D_{11}a_1 + D_{1m }a_m = \lambda^{(1)} a_1\qquad m\geq 2\\
    &D_{nm}a_m = \lambda^{(1)} a_n \qquad n\geq 2
    \end{aligned}
\end{equation}

 The first and third equations can be rewritten as
 \begin{equation}
 \begin{aligned}
     &a_1=\frac{(\Gamma_2-\lambda^{(1)})a_0}{ig-\frac{\beta\delta x}{2\sqrt{2M\omega}}} \, ,\\
     &a_n= (\lambda^{(1)}\mathbf{1}-D)^{-1}_{nm}D_{m1}a_1, \qquad n,m\geq 2
     \end{aligned}
 \end{equation}
Inserting these two relations in the second equation of \eqref{eq:hybrid_system_1}, and assuming $a_0\neq 0$, we obtain the equation
\begin{equation}
\label{eq:hybrid_syst_quad}
    (\lambda^{(1)})^2-\lambda^{(1)}(\Gamma_2+\gamma(\lambda^{(1)}))-g^2-\frac{\beta^2\delta x^2}{8M \omega}+\Gamma_2\gamma(\lambda^{(1)})=0\, ,
\end{equation}
where $\gamma(\lambda^{(1)})=D_{11}+D_{1n}(\lambda^{(1)}\mathbf{1}-D)^{-1}_{nm}D_{m1}$, with $n,m\geq 2$. Equation \eqref{eq:hybrid_syst_quad} can be implicitly solved for $\lambda^{(1)}$:
\begin{equation}
    \label{eq:hybrid_quad_sol}\lambda^{(1)}=\frac{\Gamma_2+\gamma(\lambda^{(1)})\pm \sqrt{(\Gamma_2-\gamma(\lambda^{(1)}))^2+4g^2+\frac{\beta^2\delta x^2}{2M\omega}}}{2}
\end{equation}
Our objective is determining a physical condition for which $\lambda^{(1)}<0$, guaranteeing that the system is entangled.
To do this, first observe that from the definition \eqref{eq:hybrid_pert_osc}, $D_{nm}$ is a positive semidefinite matrix. This means that if $\lambda^{(1)}<0$, then $D_{1n}(\lambda^{(1)}\mathbf{1}-D)^{-1}_{mn}D_{m1}<0$. In particular, $D_{11}>\gamma(\lambda^{(1)})$. 

Then, using \eqref{eq:hybrid_diss_app}, we can compute the evolution of the excess variance of the momentum at initial time:
\begin{equation}
    \frac{d\braket{p_1^2}_e}{dt}\Bigg|_{t=0}=\int d^3\mb k \, k_x^2f_1(k)
\end{equation}
On the other hand, using \eqref{eq:hybrid_pert_osc}, we have
\begin{equation}
    D_{11}=\int d^3\mb k \, \frac{k_x^2 f_1(k)}{2M\omega}e^{-\frac{k_x^2}{2M\omega}}\, ,
\end{equation}
so in particular
\begin{equation}
\label{eq:hybrid_ineq_chain_1}
    \frac{1}{2M\omega}\frac{d\braket{p_1^2}_e}{dt}\Bigg|_{t=0}>D_{11}>\gamma(\lambda)
\end{equation}
Also, using similar techniques to those of Appendix \ref{app:two_qubits}, we find that the rate of change of the visibility coming from the dissipative part of the dynamics
\begin{equation}
    \label{eq:hybrid_vis_decrease}\frac{d|\braket{\sigma_-}|_e}{dt}\Bigg|_{t=0}=-2\Gamma_2|\angbracket{\sigma_-(0)}_e|=-\Gamma_2 \, ,
\end{equation}
where we have used the fact that $\braket{\sigma_-(0)}=1/2$ using the initial state \eqref{eq:hybrid_initial_state}. Putting results \eqref{eq:hybrid_ineq_chain_1} and \eqref{eq:hybrid_vis_decrease} together, we have the inequality
\begin{equation}
    \frac{1}{2M\omega}\frac{d\braket{p_1^2}_e}{dt}\Bigg|_{t=0}+\left|\frac{d|\braket{\sigma_-}_e|}{dt}\right|_{t=0}>\Gamma_2+\gamma(\lambda) \, .
\end{equation}
Therefore, if this combination of noise effects is bounded by
\begin{equation}
    2g>\frac{1}{2M\omega}\frac{d\braket{p_1^2}_e}{dt}\Bigg|_{t=0}+\left|\frac{d|\braket{\sigma_-}|}{dt}\right|_{t=0}
\end{equation}
then certainly $\Gamma_2+\gamma(\lambda^{(1)})<2g$, so in particular the solution for $\lambda^{(1)}$ with the negative sign in \eqref{eq:hybrid_quad_sol} is negative, implying entanglement generation.

\section{Computations for specific models}
\label{app:models}
\subsection{Classical quantum gravity}
\label{app:cq_gravity}
The Lindblad equation of the model in the non-relativistic limit is \cite{Carney:2024izr}: 
\begin{equation}
\label{eq:Opp_Lind}
    \dot \rho=-i[H,\rho]+\int d^3\mb x\, d^3 \mb y \, F_2(\mb x-\mb y)\left[J(\mb x)\rho J(\mb y)-\frac{1}{2}\{J(\mb x)J(\mb y),\rho \}\right] \, ,
\end{equation}
where 
\begin{equation}
    H=\sum_{a=1,2}\frac{\mb p_a^2}{2m_a}-\frac{G_N m_1m_2}{|\mb x_1-\mb x_2|} \, ,
\end{equation}
and $J(\mb x)=\sum_{i=1,2}\lambda_i\delta^3(\mb x-\mb x_i)$, $F_2(\mb x-\mb y)=D_2G_C(\mb x-\mb y)+D_0\delta^3(\mb x-\mb y)$, with
\begin{equation}
    G_C(\mb x)=\int \frac{d^3\mb k}{(2\pi)^3}\frac{e^{i\mb k\cdot \mb x}}{(\mb k^2+m_\phi^2)^2} \ ,\qquad \delta^3(\mb x)=\int \frac{d^3\mb k}{(2\pi)^3}e^{i\mb k\cdot \mb x}
\end{equation}

In the above, $m_\phi $ is an infrared regulator to be sent to $0$ at the end of the computation. The coupling constants are $\lambda_a=\sqrt{4\pi G_N}m_a$. 
Using the definitions for $J(\mb x)$ and $F_2(\mb x-\mb y)$ and performing the spatial integrals, the dissipation can be written as
\begin{equation}
\label{eq:Opp_diss_full}
\begin{aligned}
    \mathcal{D}(\rho)
    =&\int d^3\mb k  \, F_2(\mb k)\Bigg[\sum_{i=1,2}\lambda_i^2 \left(e^{i\mb k\cdot \mb x_i}\rho^{-i\mb k\cdot \mb x_i}-\rho \right)+\lambda_1\lambda_2 \left(e^{i\mb k\cdot \mb x_1}\rho e^{-i\mb k\cdot \mb x_2}+e^{i\mb k\cdot \mb x_2}\rho e^{-i\mb k\cdot \mb x_1}-e^{i\mb k\cdot (\mb x_1-\mb x_2)}\rho-\rho e^{i\mb k\cdot (\mb x_1-\mb x_2)}\right)\Bigg]
    \end{aligned}
\end{equation}
where
\begin{equation}
    F_2(\mb k)=\frac{1}{(2\pi)^3}\left[\frac{D_2}{(\mb k^2+m^2_{\phi})^2}+D_0\right] \, .
\end{equation}
Restricting to $1$ spatial dimension and recurring to our small displacement approximation, \eqref{eq:1dapporx_app}, we find that the dissipation structure is of the form \eqref{eq:lind_restr}, with 
\begin{equation}
\begin{aligned}
    f_a=&\frac{G_N m_a^2}{2\pi^2}\left[\frac{D_2}{(\mb k^2+m^2_{\phi})^2}+D_0\right] \,  \, ,\\
    \beta=& 4\pi G_N m_1m_2\int d^3\mb k  \frac{e^{i\mb k_x d}}{(2\pi)^3}\left[\frac{D_2}{( k^2+m^2_{\phi})^2}+D_0\right] \approx -\frac{2G_Nm_1m_2(D_0+D_2\ell^4)\cos(d/\ell)}{\pi\ell^3 d^2}
    \end{aligned}
\end{equation}

\subsection{Entropic gravity}
\label{appendix-entropic}

For both the local and non-local models, the Lindblad equation is of the form
\begin{equation}
\label{eq:entropic_lind_app}
    \dot{\rho} =-i[H,\rho] +\sum_{\alpha,\pm}\left[K_{\alpha,\pm}\rho K_{\alpha,\pm}^\dagger-\frac{1}{2}\{K_{\alpha,\pm}^\dagger K_{\alpha,\pm},\rho\}\right] \, ,
\end{equation}
where the $K_{\alpha,\pm}$ are functions of free parameters of the model, which depend on the properties on the mediator system and the bath that regulates the temperature of such system. For details on the specific construction of these models, refer to \cite{Carney:2025wmz}.

\textbf{Non-local model}. The expressions for the $K_{\alpha,\pm}$ operators are given by
 \begin{equation}
  \label{eq:K_non_loc}
  \begin{aligned}
      K_{\alpha,+}=\sqrt{2\zeta T}\int_0^{\omega_\alpha(\mb x)/T}d\nu \, \sqrt{g_+(\nu)} \, ,\\
      K_{\alpha,-}=\sqrt{\frac{2\zeta}{ T}}\int_0^{\omega_\alpha(\mb x)/T}d\nu \, \sqrt{g_-(\nu)} \, ,
      \end{aligned}
  \end{equation}
where $\alpha=1,2,3,\dots $ is an index which labels the mediators and $\mb x$ denotes the relative position $\mb x=\mb x_1-\mb x_2$.  Here, $T$ is the temperature of the mediator state and $\zeta$ is a parameter that controls the mediator damping rate. The expressions for $g_\pm (\nu)$ are

\begin{equation}
\begin{aligned}
g_{+}(\nu) =& \frac{1}{64\nu} \csch^{3}(\nu/2) \sech(\nu/2) \left[ -2 + (2 + \nu^{2}) \cosh(\nu) - 2\nu \sinh(\nu) \right] \, ,\\
g_{-}(\nu )=&\frac{2}{\nu}\csch^3(\nu)\sinh^4(\nu/2) \, .
\end{aligned}
\end{equation}
The functions $\omega_\alpha(\mb x)$ describe the frequencies of the various modes of the mediator:
  \begin{equation}
  \label{eq:omega_non_loc}
      \omega_\alpha(\mb x)=\alpha f(\mb x)\,, \qquad f(\mb x)=\left(\lambda + \frac{\ell^2}{|\mb x_1-\mb x_2|}\right)^{-1}\,,
  \end{equation}
  with $\lambda,\ell^2$ are free parameters with dimensions of length. To recover the Newton force, the parameter $\ell^2$ is further subject to the constraint
  \begin{equation}
  \label{eq:emergent_G_non_loc}
      \frac{\pi^2}{12}T^2\ell^2 = G_N m_1m_2 \, .
  \end{equation}
  Since the Lindblad operators only depend on the relative distance, for experimental setups involving two masses at a large initial separation $d$, the Gaussian approximation is valid, and the dissipation assumes the form   \begin{equation}
  \mathcal{D(\rho)}=\frac{1}{2}\sum_{\alpha,\pm} (\partial_rK_{\alpha,\pm}|_{r=d})^2[r,[\rho,r]]=\frac{\beta}{2}[r,[\rho,r]] \, ,
  \end{equation}
  To compute the spatial derivative of the $K_{\alpha,\pm}$ operators, we apply the chain rule and then substitute $\omega_{\alpha}(\mb x)=\alpha f(\mb x)$ into the upper bound of integration. Then, we take the continuum limit $\sum_\alpha\rightarrow \int_0^\infty d\omega/f(\mb x)$, as outlined in \cite{Carney:2025wmz}. The result is
  \begin{equation}
      \beta= \frac{G_Nm_1m_2}{L^3(1+\lambda L/\ell^2)}\left(\zeta\mathcal{I}_++\frac{\mathcal{I}_-}{\zeta}\right)
  \end{equation}
  where $\mathcal{I_{\pm}}$ are numerical values obtained by the following integrals of functions $g_{\pm}(\nu)$:
  \begin{equation}
  \begin{aligned}
      \mathcal{I_+}=\int_0^\infty d\nu \, \nu^2 g_{+}(\nu)\approx 1.17 \, ,&\\
      \mathcal{I_-}=\int_0^\infty d\nu \, \nu^2 g_{-}(\nu)\approx1.21 \, .&
      \end{aligned}
  \end{equation}
  As discussed in \ref{sec:examples}, the expression for $\beta$ corresponds to the force noise predicted by the model, in an experiment involving two masses $m_1,m_2$ placed at a distance $d$.

\textbf{Local model}. The expressions for the $K_{\alpha,\pm}$ operators are given by
 \begin{equation}
     K_{\alpha,\pm}=\lambda_{\pm}\omega_{\alpha}(\mb x) \, ,
 \end{equation}
with 
\begin{equation}
    \lambda_+=\sqrt{\frac{\sigma_*\gamma_{th}}{4T^2}} \, , \qquad \lambda_-=\sqrt{\frac{2\sigma_*(\sigma_*-1)^2}{\gamma_{th}}}\ , 
\end{equation}
where $T$ is the equilibrium temperature of the mediator, $\sigma_*$ is the average polarization of the spins, and $\gamma_{th}$ is the thermalization rate. Here $\alpha$ is a discrete index that labels lattice sites. The functions $\omega_{\alpha}$ are
 \begin{equation}
     \omega_\alpha(\mb x)=\sum_a \frac{\ell_a}{|\mb r_\alpha-\mb x_I|^2+a^2} \, ,
 \end{equation}
where $\mb r_\alpha$ are the lattice coordinates, $a$ is the lattice scale, and $\ell_a=m_a L^2$, with $L$ being a free length scale of the model.
 To recover the emergent Newtonian force, Newton's constant is defined in terms of some of the constants introduced above:
 \begin{equation}
     \frac{\sigma_*(1-\sigma_*)}{T}\frac{\pi^3L^4}{a^3}\equiv G_N
 \end{equation}
Taking the continuum limit $\sum_{\alpha}\rightarrow\int d^3r/a^3$, and using the Fourier transform formula
\begin{equation}
    \frac{1}{|\mb r-\mb x_a|^2+a^2}=\int\frac{d^3\mb k}{(2\pi)^3}\frac{2\pi^2}{k} e^{-ak}e^{i\mb k\cdot (\mb r-\mb x_a)} \, ,
\end{equation}
a straightforward calculation shows that the lindblad equation \eqref{eq:entropic_lind_app} for the local model reduces to our Eq. \eqref{eq:lind_restr}, with 
 \begin{equation}
 \begin{aligned}
     &f_a(\mb k)=\frac{\pi (\lambda_+^2+\lambda_-^2)m^2_aL^4}{2a^3}\frac{e^{-2ak}}{k^2}\\
     &\beta=\frac{4\pi^2}{a^3}(\lambda_+^2+\lambda_-^2)m_1m_2L^4\left[\frac{4a(2a^2+d^2)}{4a^2d+d^3}-\frac{\arctan(d/2a)}{d^3}\right]\\
     &\approx-\frac{2\pi^3}{a^3d^3}(\lambda_+^2+\lambda_-^2)m_1m_2L^4 \, , 
     \end{aligned}
 \end{equation}
 where in the last line we have assumed that $d\gg a$.
\end{appendix}

\end{document}